%Paper: hep-th/9407080
%From: nietol@ERE.UMontreal.CA (Nieto Luis M.)
%Date: Fri, 15 Jul 94 01:24:44 -0400
%Date (revised): Fri, 15 Jul 94 11:01:39 -0400

%%%%%%%%%%%%%%%%%%%     AMS-TEX   %%%%%%%%%%%%%%%%%
\magnification\magstep1
\documentstyle{amsppt}
\TagsOnRight
\hsize 6.3truein
\vsize 8.5truein
\baselineskip=12pt
\NoBlackBoxes
%\printoptions
%\font\ninepoint=cmr9
\font\titulo=cmbx10 scaled\magstep2
%\NoPageNumbers
\font\titulo=cmbx10 scaled\magstep2
\font\autor=cmbx10 scaled \magstep1
%%%%%%%%%%%%%%%%%%%%%%%%%%%%%%%%%%%%%%%%%%%
\define\h{ {\Cal H}^4}
\define\ph{ P{\Cal H}^4}
\define\es{SU(2,2)/SU(1,2)}
\define\g{g_{\mu\nu}}
\define\ho{O(2,2)/O(2,1)}

%%%%%%%%%%%%%%%%%%%%%%%%%%%%%%%%%%%%%%%%%%%%

\centerline{\titulo THE CONFORMAL GROUP SU(2,2)}
\bigskip
\centerline{\titulo AND INTEGRABLE SYSTEMS}
\bigskip
\centerline{\titulo  ON A LORENTZIAN HYPERBOLOID}
\vskip 2truecm
\centerline{\autor M.A. del Olmo$^1$, M.A. Rodr\'{\i}guez$^2$
and P.  Winternitz$^3$}
\bigskip \centerline {$^1$Dept. F\'{\i}sica Te\'orica,
Facultad de  Ciencias,}
\centerline {Universidad de Valladolid, E-47011 Valladolid,
(Spain).}
\smallskip
\centerline {$^2$Dept. F\'{\i}sica Te\'orica, Facultad de
F\'{\i}sicas,}
\centerline{Universidad Complutense, E-28040 Madrid (Spain).}
\smallskip
\centerline{$^3$Centre de recherches math\'ematiques,
Universit\'e de
 Montr\'eal}
\centerline {CP 6128, Succ. Centre--Ville, Montr\'eal,  H3C
3J7, Qu\'ebec (Canada).}

\vskip 2cm
\centerline {June 23, 1994}
\vskip 1cm
\centerline {CRM $\#$ 2194}
\vskip 1cm
Runing title: ``$SU(2,2)$--INTEGRABLE SYSTEMS"
\bigskip PACS $\#$: 02.20.Sv; 02.30.Jr; 03.20.+i; 03.65.Fd;
11.30.Cp
\bigskip\bigskip
\noindent {\eightpoint {\bf Abstract.}
Eleven different types of ``maximally superintegrable"
Hamiltonian  systems on the real hyperboloid
$(s^0)^2-(s^1)^2+(s^2)^2-(s^3)^2=1$ are  obtained. All of
them correspond to a free  Hamiltonian system on  the
homogeneous space $SU(2,2)/U(2,1)$, but to reductions by
different  maximal abelian subgroups of $SU(2,2)$. Each of
the obtained  systems  allows $5$ functionally independent
integrals of motion, from   which it is possible to form two
or more triplets in involution  (each of them includes the
hamiltonian). The corresponding classical and quantum
equations of motion can be solved  by separation of
variables on  the $O(2,2)$ space.}
\vfill\eject

%%%%%%%%%%%%%%%%%%%%%%%%%%%--INTRODUCTION--%%%%%%%%%%%%%%%%
\centerline{\bf 1.- INTRODUCTION.} \bigskip
	In a recent article [1] we constructed a family of
completely integrable finite-dimensional Hamiltonian systems
on a real $O(p,q)$ hyperboloid
$$
g_{\mu\nu}s^{\mu}s^{\nu}=1,\qquad  g=g^T \in  {\Bbb
R}^{(p+q)\times (p+q)},
 \tag 1.1
$$
$$
\text{sign}(g)=(p,q), \quad p \geq q \geq 0, \quad s \in
\Bbb R^{p+q}.
$$
The classical Hamiltonian had the form
$$
H=\frac{1}{2}g^{\mu\nu}p_{s^\mu}p_{s^\nu}+V(s), \tag 1.2a
$$
where $p_{\mu}$ are the momenta canonically conjugate to
the coordinates
$s^{\mu}$  and the corresponding quantum mechanical one is
$$
H=-\frac{1}{2}\sum ^n
_{\mu ,\nu =1}\frac 1 {\sqrt {g_0}} \frac {\partial}
{\partial s^{\mu }}
\sqrt{g_0}\ g ^{\mu \nu }\frac {\partial} {\partial   s^{\nu
}}+V(s),
\tag1.2b
$$
with $g_0= \det g$. The momenta $p_{s^\mu}$ satisfy
$$
p_{s^\mu }s^{\mu}=0. \tag1.3
$$
\smallskip
The potential $V(s)$ was obtained by projecting free  motion
on a projective complex hyperboloid
$$
P{\Cal H}^{n+1} \sim SU(p,q)/U(p-1,q), \quad p+q=n+1,  \tag
1.4
$$
onto the real $O(p,q)$ hyperboloid.
\smallskip
The starting point is thus a free Hamiltonian
$$
H_F=\frac{1}{2}g^{\mu\nu}\bar{p}_{\mu}p_{\nu}, \tag 1.5
$$
where $p_{\mu}$ are the complex momenta canonically
conjugate to the  complex coordinates $y^{\mu}$, satisfying
$$
g_{\mu\nu}\bar{y}^{\mu}y^{\nu}=1. \tag 1.6
$$
The projection was performed by introducing $n$ ignorable
variables on
$P{\Cal H}^{n+1}$, corresponding to the diagonalization of
all elements of  a Cartan subalgebra of $u(p,q)$. We recall
that a variable, in a certain coordinate system, is called
ignorable if the corresponding metric tensor
$g$ does not depend on it. Since $u(p,q)$ (for$p \geq q$)
has $q+1$  different Cartan subalgebras $M_r,\, r=0,1,...,q,$
we obtained $q+1$ different integrable hamiltonian systems.
\smallskip
In addition to Cartan subalgebras, that are maximal abelian
and selfnormalizing subalgebras [2], the $su(p,q)$ algebras
have maximal  abelian subalgebras (MASAs) that contain
nilpotent elements. All MASAs of
$su(p,q)$  have recently been classified [3] and in
particular $su(2,2)$ has 12
$SU(2,2)$--conjugacy of MASAs, 3 of them Cartan subalgebras.
\smallskip
	The aim of this article is to show that non Cartan MASAs
can also be used to introduce ignorable variables and to
generate integrable systems of the type (1.2).
\smallskip
We shall restrict our study to the algebra $su(2,2)$ and
to the homogeneous (and symmetric) space $SU(2,2)/U(2,1)$.
This is actually  a case of particular physical interest,
since $SU(2,2)$ is locally isomorphic  to the conformal
group $C(3,1)$ of Minkowski space-time. We are thus studying
conformally invariant integrable systems on the space (1.4).
The integrability of the systems with Hamiltonian (1.2) is a
consequence of the original conformal invariance of the free
Hamiltonian system on
$P{\Cal H}^4$.
\smallskip
The motivation for generating new integrable systems was
discussed in our previous article [1]. We mention here that
the Hamiltonian systems obtained in this article are not
just integrable, but rather ``maximally superintegrable"
[4,5]. This means that instead of allowing $n$ integrals of
motion in involution (for $n$ degrees of freedom), such
systems allow $2n-1$ functionally independent integrals of
motion that are well defined functions on phase space.
Amongst these, different sets of $n$ integrals in involution
can be chosen. Such superintegrable systems are rare. They
include the Coulomb system, the harmonic oscillator and a few
others [5,\dots,12]. The classical trajectories in such
systems are periodic, if  they are finite. The energy levels
of the corresponding quantum systems are degenerate. The
integrals of motion generate a Lie algebra, the
representation theory of which explains the ``accidental"
degeneracy [5,\dots,12]. The underlying group structure
makes it possible to solve the equations of motion
analytically: a very usefull feature in applications.
\smallskip
Completely integrable potentials such as the P\"oschl-Teller
potential [13],  have found interesting applications in
molecular, atomic, nuclear and  particle physics
[14,\dots,17].
\bigskip\bigskip
%%%%%%%%%%--CAPITULO II--%%%%%%%%%%%%%%%%%%
\centerline {\bf 2.- THE GROUP SU(2,2) AND ITS LIE ALGEBRA.}
\bigskip
A short review of some pertinent information on the group
$SU(2,2)$, its  Lie algebra $su(2,2)$ and the homogeneous
space $P{\Cal H}^4$ will be presented  in  this section.
Most of its contents are classical results on complex
geometry and Lie theory.
\bigskip
\noindent
{\bf 2.1.- The su(2,2) Lie Algebra.}
\medskip
We shall realize the Lie algebra $su(2,2)$ and group
$SU(2,2)$ by matrices
$X$ and $G$ satisfying
$$ \alignat 3
XK+KX^\dagger &=0, &\qquad \text{Tr}X=0, &\qquad K, \ X,\ G
\in
\Bbb C^{4\times 4}  \tag 2.1
\\ GKG^\dagger &=K, &\qquad \text{det}G=1, &\qquad
K=K^\dagger,
\endalignat
$$
where $K$ is a hermitian matrix of signature $(++--)$.
\smallskip
A convenient basis for $su(2,2)$ is provided by 15 matrices
$X_k, k=1,\dots,15$.  Their specific form depends on the
choice of the  matrix
$K$. \smallskip
We shall actually need 6 different bases, corresponding to
6 realizations  of
$su(2,2)$ (with different matrices $K$). The transformation
from one  realization to another is given by
$$
gKg^+ =K^\prime, \qquad  gXg^{-1} =X^{\prime}, \qquad g \in
SL(4,\Bbb C),  \tag 2.2
$$
$$
X_{(1)} =\left[\matrix
i w_{1}&w_{4}+i w_{5}&w_{6}+i w_{7}&w_{8}+i w_{9}\\
w_{4}-i w_{5}&-i w_{1}+i w_{2}&w_{10}+i w_{11}&w_{12}+i
w_{13}\\ -w_{6}+i w_{7}&w_{10}-i w_{11}&-i w_{2}+i
w_{3}&w_{14}+i w_{15}\\ w_{8}-i w_{9}&-w_{12}+i
w_{13}&w_{14}-i w_{15}&-i w_{3}
\endmatrix\right],
K_1 =\left[\matrix
1&0&0&0\\ 0&-1&0&0\\ 0&0&1&0\\ 0&0&0&-1
\endmatrix\right],\tag 2.3a
$$
$$
X_{(2)} =\left[\matrix
i w_{1}+i w_{2}&w_{4}+i w_{5}&w_{6}+i w_{7}&w_{8}+i w_{9}\\
w_{4}-i w_{5}&-i w_{1}+i w_{2}&w_{10}+i w_{11}&w_{12}+i
w_{13}\\ -w_{8}+i w_{9}&w_{12}-i w_{13}&-i w_{2}+w_{3}&i
w_{14}\\ -w_{6}+i w_{7}&w_{10}-i w_{11}&i w_{15}&-i
w_{2}-w_{3}
\endmatrix\right],
K_2 =\left[\matrix
1&0&0&0\\ 0&-1&0&0\\ 0&0&0&1\\ 0&0&1&0
\endmatrix\right], \tag 2.3b
$$
$$
X_{(3)} =\left[\matrix
w_{1}+i w_{3}&i w_{4}&w_{6}+i w_{7}&w_{8}+i w_{9}\\
   i w_{5}&-w_{1}+i w_{3}&w_{10}+i w_{11}&w_{12}+i w_{13}\\
   -w_{12}+i w_{13}&-w_{8}+i w_{9}&w_{2}-i w_{3}&i w_{14}\\
   -w_{10}+i w_{11}&-w_{6}+i w_{7}&i w_{15}&-w_{2}-i w_{3}
\endmatrix\right],
 K_3 =\left[\matrix
0&1&0&0\\ 1&0&0&0\\ 0&0&0&1\\ 0&0&1&0
\endmatrix\right],\tag 2.3c
$$
$$
X_{(4)} =\left[\matrix
w_{1}+i w_{2}&w_{6}+i w_{7}&i w_{4}&w_{10}+i w_{11}\\
   w_{8}+i w_{9}&-i w_{2}+i w_{3}&-w_{6}+i w_{7}&w_{12}+i
w_{13}\\
   i w_{5}&-w_{8}+i w_{9}&-w_{1}+i w_{2}&w_{14}+i w_{15}\\
   w_{14}-i w_{15}&w_{12}-i w_{13}&w_{10}-i w_{11}&-i
w_{2}-i w_{3}
\endmatrix\right],
K_4 =\varepsilon \left[\matrix
0&0&1&0\\ 0&1&0&0\\ 1&0&0&0\\ 0&0&0&-1
\endmatrix\right], \tag 2.3d
$$
$$
X_{(5)} =\left[\matrix
w_{1}+i w_{2}&w_{4}+i w_{5}&i w_{8}&w_{10}+i w_{11}\\
   w_{6}+i w_{7}&w_{3}-i w_{2}&-w_{10}+i w_{11}&i w_{9}\\
   i w_{12}&w_{14}+i w_{15}&-w_{1}+i w_{2}&-w_{6}+i w_{7}\\
   -w_{14}+i w_{15}&i w_{13}&-w_{4}+i w_{5}&-w_{3}-i w_{2}
\endmatrix\right],
K_5 =\left[\matrix
0&0&1&0\\ 0&0&0&1\\ 1&0&0&0\\ 0&1&0&0
\endmatrix\right], \tag 2.3e
$$
$$
X_{(6)} =\left[\matrix
w_{1}+i w_{2}&w_{4}+i w_{5}&w_{6}+i w_{7}&i w_{12}\\
   w_{8}+i w_{9}&w_{3}-i w_{2}&i w_{13}&-w_{6}+i w_{7}\\
   w_{10}+i w_{11}&i w_{14}&-w_{3}-i w_{2}&-w_{4}+i w_{5}\\
   i w_{15}&-w_{10}+i w_{11}&-w_{8}+i w_{9}&-w_{1}+i w_{2}
\endmatrix\right],
K_6 =\left[\matrix
0&0&0&1\\ 0&0&1&0\\ 0&1&0&0\\ 1&0&0&0
\endmatrix\right], \tag 2.3f
$$
where $\varepsilon = \pm 1.$
\smallskip
We denote the matrix elements of $X_{(a)}, \, (a=1,\dots,6)$,
$w_i \,  (i=1,\dots,15)$ in all cases,
though a  notation $\omega_{(a)i}$ would be more consistent.
Notice that a  basis element $X_i$, obtained by setting
$\omega_i=1, \, \omega_j =0, \, j\neq i$, in one realization
is not necessarily conjugate to $X_i$ in another realization.
For instance $X_3$  in the realization of $K_1$ generates a
rotation (compact), in the realization of $K_2$ the element
$X_3$ corresponds to a hyperbolic rotation  (non compact).
Similarly, $X_8$ in the case of $K_1$ corresponds to a
hyperbolic  rotation, whereas $X_8$ in the case of $K_5$ is
nilpotent.
\smallskip The form of the second order Casimir operator
$C_2$ of $su(2,2)$ is basis  dependent. For instance in the
first case, with $K_1$ adapted to the Cartan subalgebras, we
have
$$ \split
C  =
& \, 3X_1^2+4X_2^2+3X_3^2+2\{X_1,X_2\}+2\{X_2,X_3\}+ \{X_1,
X_3\} \\ &\quad
+2(-X_4^2-X_5^2+X_6^2+X_7^2-X_8^2-X_9^2-X_{10}^2- X_{11}^2 \\
& \qquad +X_{12}^2+X_{13}^2-X_{14}^2-X_{15}^2).   \endsplit
\tag 2.4
$$
\bigskip
\noindent
{\bf 2.2.- The Homogeneous Space SU(2,2)/U(1,2).}
\medskip
In Ref.1 we have constructed the space $SU(p,q)/U(p-1,q)$
in general, so we shall not present any details here. The
group $SU(2,2)$ acts transitively  on the complex
hyperboloid
$\h \sim \es $,
$$
y^+Ky=1. \tag 2.5
$$
If we identify any two points $y$ and $y^\prime$ on $\h$,
satisfying
$$
y^\prime=ye^{i\phi}, \qquad 0 \leq \phi < 2\pi, \tag 2.6
$$
we obtain the projective hyperboloid
$$
\ph \sim SU(2,2)/U(1,2). \tag 2.7
$$
Its real dimension is 6. The identification (2.6) can be
realized by introducing affine coordinates
$$
z^j=\frac{y^j}{y^0}, \qquad j=1,2,3. \tag 2.8
$$
The flat Hermitian metric
$$
h(y,y^\prime)=\g y^\mu\bar{y}^{\prime \nu} \tag 2.9
$$
then reduces to a noncompact version [1] of the Fubini-Study
metric [18].
\medskip
Real vector fields in the tangent space $T\h$ to the
hyperboloid $\h$,  realizing the Lie algebra $u(2,2)$,
acting on differentiable functions on
$\h$, are given by
$$
\hat{Z}=-y^\mu (Z)^\nu_\mu\partial_{y^\nu}+c.c., \quad Z
\in u(2,2),
\tag 2.10
$$
where c.c. denotes complex conjugation. We include the
$u$(1) basis element for which the vector field is
$$
\hat Y_0= -i(y^0 \partial_{y^0} + y^1 \partial_{y^1} +
          y^2 \partial_{y^2} + y^3 \partial_{y^3}).   \tag
2.11
$$
\smallskip
Functions $F(y)$ on $\h$ that project properly onto $\ph$
must be  homogeneous, i.e. satisfy
$$
F(y^0,y^1,y^2,y^3)=F(1,y^1/y^0,y^2/y^0,y^3/y^0). \tag 2.12
$$
On these functions we have
$$
\hat Y_0F=0, \tag 2.13
$$
and for the corresponding constant of motion we get
$$
y^0p_0 +y^1p_1 + y^2p_2 + y^3p_3=0.
$$
\bigskip
\noindent
{\bf 2.3.- Maximal Abelian Subalgebras of su(2,2).}
\medskip
Below in Section 3 we shall separate variables in the
Hamilton--Jacobi, or the Laplace--Beltrami equation on the
homogeneous space  $\ph \sim SU(2,2)/U(1,2)$. To do this we
need to construct  complete sets of commuting first and
second order operators in the enveloping algebra of
$su(2,2)$. The coordinate systems relevant for us will
involve three ignorable variables, i.e. coordinates that do
not occur in the corresponding metric tensor [19]. Such
coordinates are associated with MASAs of the Lie algebra of
the isometry group of the considered space [20], in our case
$SU(2,2)$. The ignorable coordinates are obtained by
simultaneously diagonalizing all elements of a MASA. This is
possible, if the vector fields representing the action of the
MASA on differentiable functions over the considered space,
are linearly independent at generic points of the space.
\medskip
The MASAs of $su(2,2)$ were recently classified as an
application of a  general study of MASAs of $su(p,q)$ for $p
\geq q \geq 1$ [3]. We reproduce the results  in Table 1.
There are 12 $SU(2,2)$ conjugacy classes of MASAs  of
$su(2,2)$.  Their type is given in column 2. The first 7
MASAs are orthogonally decomposable  and we give the
decomposition pattern.  The first 3 of them are Cartan
subalgebras, containing maximal compact subalgebras of
dimension 3, 2 and 1,  respectively. The MASA $M_9$ is
decomposable, but not orthogonally.  The MASAs $M_{10},\,
M_{11}$ and
$M_{12}$ are indecomposable and they are  maximal abelian
nilpotent algebras (the matrices representing them are
nilpotent). In Column 3 we give a basis for each MASA. The
corresponding  $su(2,2)$ matrices can be read off from one of
equations (2.3). Which one  is to be used is indicated by the
form of the metric given in Column 4. \smallskip	 The choice
of representative MASAs in Table 1 is not the same as in Ref.
3. The reason is that it is convenient, as will become clear
below, to choose representatives that are pure imaginary
matrices $X=-\bar{X}$. That is possible for $M_{1}, \dots,
M_{11}$,  not however  for $M_{12}$.
\smallskip
In all cases, if we add the basis element $iI$ to
$M_a, \, 1 \leq a \leq 12$,  we obtain a MASA of $u(2,2)$.
\vfill\eject
%%%%%%%%%%%%%%%%%--CAPITULO III%%%%%%%%%%%%%%%%%%
%\bigskip\bigskip
\centerline {\bf 3.- IGNORABLE COORDINATES ON $P{\Cal
H}^{n+1}$ AND}
\smallskip
\centerline{\bf THE REDUCTION TO AN O(p,q) HYPERBOLOID.}
\bigskip

While this article is essentially devoted to the
six-dimensional space
$SU(2,2)/U(1,2)$ we shall present results in this section
for arbitrary
$p$ and $q$ satisfying $p \geq q \geq 0, \, p+q=n+1$.
\bigskip
\noindent
{\bf 3.1.- The Ignorable Variables.}
\medskip
In Ref.1 we developed a general method for introducing
coordinates on the space $P{\Cal H}^{n+1}$ containing $n+1$
ignorable variables, corresponding  to the elements of a
Cartan subalgebra. Here we shall generalize these  results
to a larger class of abelian subalgebras of $u(p,q)$.
\smallskip
\proclaim{Theorem 1} Let M be an abelian subalgebra of
$u(p,q),\, p \geq q \geq 0, \, p+q=n+1$, satisfying the
following conditions:
\roster
\item   $\text{dim} M= p+q=n+1$.
\item  The $n+1$ vectors fields
$$
\hat Y_a= -y^\mu(Y_a)^\nu _\mu \partial_{y^\nu}, \quad a=0,1,
\dots, n,
\tag 3.1
$$
(where $Y_a$ are the $u(p,q)$ matrices representing the
abelian algebra
$M$)  are linearly independent at a generic point of
$P{\Cal H}^{n+1}$.

\item All elements $Y_a$ are represented by pure imaginary
matrices in the considered realization in $\Bbb C^{(p+q)
\times (p+q)}$.
\endroster
Then the vector fields $\hat{Y}_a+\bar{\hat{Y}}_a,\, a=0,1,
\dots, n$  representing  the algebra $M$ in the tangent space
$T{\Cal H}^{n+1}$ can be  simultaneously  straightened out to
$$
\hat Y_\mu+\bar{\hat{Y}}_{\mu}= -\partial_{x^\mu}, \quad
\mu=0,1, \dots, n. \tag 3.2 $$
This is achieved by a transformation of coordinates
$(y^\mu,\ \bar{y}^{\mu}) \to (s^\mu,\ x^{\mu})$, where
$s^\mu$ and
$x^\mu$ are real and satisfy
$$
y^\mu=B(x)^\mu _\nu s^\nu, \qquad B(x)= \exp(x^\rho Y_\rho).
\tag 3.3
$$
The matrices $Y_\rho$ span the considered MASA, the
variables $x^\rho$  are the ignorable ones, while $s$
satisfies
$$
s^TKs=1, \tag 3.4
$$
when $y$ satisfies eq.(2.5). The vector fields $\hat Z$ of
eq.(2.10)  representing the algebra $u(p,q)$, reduce to
$$
\hat{Z}=-
\frac{1}{2}B^{\mu}_{\nu}s^{\nu}(Z)^{\alpha}_{\mu}[
(A^{-1})^\beta_\alpha
\frac{\partial}{\partial x^\beta}+ (B^{-1})^\beta_\alpha
\frac{\partial}{\partial s^\beta}] , \tag 3.5
$$
with
$$
A^\mu_\nu =\frac{\partial y^\mu}{\partial x^\nu}
  =(Y_\nu)^{\mu}_{\rho}y^{\rho}. \tag 3.6
$$
\endproclaim\smallskip
{\sl Proof}. The proof coincides with the proof of Theorem 1
of Ref.1, though the Theorem presented in this paper is more
general. We shall not repeat the  proof here. Useful elements
of it are that the Jacobian of the transformation  (3.3) and
its inverse are  $$ j=\frac{\partial (y,\bar{y})}{\partial
(x,s)}=
\left ( \matrix
A &B \\ \bar{A} & \bar{B} \endmatrix\right ), \quad
j^{-1}=\frac{1}{2}\left (
\matrix A^{-1} & \bar{A}^{-1}\\B^{-1} &
\bar{B}^{-1} \endmatrix\right ), \tag 3.7
$$
with $A$ as in (3.6) and $B$ as in (3.3). We hence  have
$$
\frac{\partial }{\partial y^\mu}=
\frac{1}{2}[(A^{-1})^\nu_\mu \frac{\partial}{\partial
x^\nu}+(B^{-1})^\nu_\mu
\frac{\partial}{\partial s^\nu}]. \tag 3.8
$$
%\vfill\eject\bigskip
\noindent
{\bf 3.2.- Reduction of the Elements of the Lie Algebra.}
\medskip
Let us consider a basis for $su(p,q)$ in which all basis
elements are either real, or pure imaginary (all bases
in (2.3) satisfy this requirement). For such a
basis eq.(3.5) simplifies. Indeed, let $X_{Re}$ and $X_{Im}$
denote real and imaginary $su(p,q)$ matrices, respectively.
The corresponding vector fields are expressed in terms of
the  coordinates $(s,x)$ and conjugate momenta $(p_s,p_x)$ as
$$ \align
\hat X_{Re} &= -p_s^T(X_{Re})s, \tag 3.9 \\
\hat X_{Im} &= -p_x^TA^{-1}(X_{Im})s. \tag 3.10
\endalign
$$

In order to restrict to the real hyperboloid of eq.(3.4),
we set $x=0$  and $p_x=\text{constant.}$ The vector fields
$\hat X_{Re}$ then generate  the algebra $o(p,q)$ and lie in
the tangent space to the $O(p,q)$  hyperboloid. Their form
depends only on that of the matrices $X_{Re}$,  i.e. of the
chosen metric. The vector fields $\hat X_{Im}$, on the  other
hand, reduce to functions of $s$, depending on some constants
$k=(k_0, k_1, \dots, k_n)$. Their form depends on $A$ and
is hence  different for each MASA.
\bigskip
\noindent
{\bf 3.3.- Reduction of the Hamiltonian.}
\medskip
The free Hamiltonian on the space $\ph$, corresponding to
geodesic  motion on this space, is
$$
H=\frac{c}{2}g^{\mu\nu}\bar{p}_{\mu}p_{\nu}, \tag 3.11
$$
where $p_\mu$ are the complex momenta canonically conjugate
to $y^\mu$.  In the coordinates $(s,x)$ the Hamiltonian
reduces to
$$
H= \frac{c}{8}[p_s^Tgp_s +p_x^T(A^{\dagger}gA)^{-1}p_x].
\tag 3.12
$$
Restricting to the $O(p,q)$ hyperboloid we obtain the
Hamiltonian (1.2)  (for $c=4$) with the potential
$$
V(s)=\frac{c}{8}p_x^T(A^{\dagger}gA)^{-1}p_x, \tag 3.13
$$
where $p_x \in \Bbb R^{n+1}$ is a constant vector.
\smallskip
Thus the ``kinetic energy" part in eq.(3.12) is always  the
same, but the potential $V(s)$ depends on the matrix $A$ and
is hence different for each specific abelian subalgebra of
$su(p,q)$ (satisfying the conditions  of Theorem 1).
%\vfill\eject
\bigskip\bigskip
%%%%%%%%%%%%%%%%%%%%%%%%--CAPITULO IV--%%%%%%%%%%%%%%%
\centerline{\bf 4.- THE COORDINATE SYSTEMS}
\smallskip
\centerline {\bf AND REDUCED HAMILTONIANS FOR SU(2,2).}
\bigskip
We now return to the case $p=q=2$ and make the results of
Section 3  concrete for each of the MASAs of Table 1. The
MASAs $M_1, M_2, \dots, M_{11}$  all satisfy the conditions
of Theorem 1, however the four-dimensional MANS
$M_{12}$ does not. Indeed the corresponding vector fields of
eq.(3.1)  for it  are
$$\alignat 2
&\hat Y_1 \equiv \hat X_6 = -(y^2 \partial_{y^0} - y^3
\partial_{y^1}),  &\quad
&\hat Y_2 \equiv \hat X_7 = -i(y^2 \partial_{y^0} + y^3
\partial_{y^1}),
\\
&\hat Y_3 \equiv \hat X_{12} = -iy^3 \partial_{y^0}, &\quad
&\hat Y_4 \equiv \hat X_{13}= -iy^2 \partial_{y^1},\\
&\hat Y_0 \equiv \hat X_0 = -i(y^0 \partial_{y^0} + y^1
\partial_{y^1} +  y^2 \partial_{y^2} + y^3 \partial_{y^3}).
& \tag 4.1
\endalignat
$$
They satisfy
$$\align
y^2y^3 \hat {Y}_1 +i(y^2)^2 \hat {Y}_{3} -  i(y^3)^2 \hat
{Y}_{4} &=0,\\
y^2y^3 \hat {Y}_2 -(y^2)^2 \hat {Y}_{3} -  (y^3)^2 \hat
{Y}_{4} &=0,  \tag 4.2
\endalign
$$
so only 3 of them are linearly independent at a generic
point of $\h$  (and we need 4).
\medskip
We shall now run through the other MASAs, $M_1, M_2, \dots,
M_{11}$  and in each case specialize eq.(3.3) and (3.13) to
obtain the coordinates  on the complex hyperboloid of
eq.(2.5) and the real potential, figuring  in eq.(1.2). The
potential $V(s)$ will in each case be given in coordinates
adapted to the specific metric $K_a,\;(a=1,\ldots,6)$ that
we use. Note  that  using formula (3.3),  the expression of
the coordinates would be
$y^\mu=(\exp(c^\rho Y_\rho)^\mu _\nu )s^\nu $, where $c^\mu$
are the  ignorable  variables. However, in order to  simplify
notation we will write
$x^\mu$ for  the ignorable variables, which will be either
the same as
$c^\mu$ or linear combinations of them (this last case can
be seen as a basis change for the MASA). The momenta $p_x$
conjugate to $x^\mu$ or linear combinations of them will be
denoted by $k$.
\vfill\eject %\bigskip
\noindent
{\bf 1.- The Compact Cartan Subalgebra $M_1$; Metric $K_1$}
\medskip \noindent
Coordinates:
$$
y^\mu=e^{ix^\mu}s^\mu, \qquad \mu=0,1,2,3. \tag 4.3
$$
Hamiltonian:
$$
H_1=\frac{1}{2}\left(p_0^2 - p_1^2 + p_2^2 - p_3^2 +
\frac{k_0^2}{(s^0)^2}-\frac{k_1^2}{(s^1)^2}+
\frac{k_2^2}{(s^2)^2}-\frac{k_3^2}{(s^3)^2}\right). \tag 4.4
$$
The potential is singular along each of the surfaces
$s^\mu=0$, i.e.  on two--sheeted two--dimensional
hyperboloids for $s^0=0$, or $s^2=0$ and one--sheeted
hyperboloids for $s^1=0$, or $s^3=0$.
%\vfill\eject
\bigskip
\noindent
{\bf 2.- The Noncompact Cartan Subalgebra $M_2$; Metric
$K_1$.}
\medskip \noindent
Coordinates:
$$\alignat 2
y^0 & =  e^{ix^0}s^0, &\quad y^1 & =  e^{ix^1}s^1,   \tag
4.5\\ y^2 & =  e^{ix^2}(s^2 \cosh x^3+is^3 \sinh x^3),
&\quad y^3 & =  e^{ix^2}( -is^2 \sinh x^3+s^3 \cosh x^3).
\endalignat
$$
Hamiltonian:
$$
H_2= \frac{1}{2}\left (p_0^2 - p_1^2 + p_2^2 - p_3^2 +
\frac{k_0^2}{(s^0)^2} -
\frac{k_1^2}{(s^1)^2}
+ \frac{((s^2)^2-(s^3)^2)( k_2^2-k_3^2)
+ 4s^2s^3k_2k_3}{((s^2)^2+(s^3)^2)^2} \right ).  \tag 4.6
$$
The potential $V_2(s)$ is singular along the hyperboloids
$s^0=0$ and
$s^1=0$ and also along the hyperbola $s^2=s^3=0, \,
(s^0)^2-(s^1)^2=1$.
\bigskip
\noindent
{\bf 3.- The Noncompact Cartan Subalgebra $M_3$; Metric
$K_1$.}
\medskip \noindent
Coordinates:
$$\alignat 2
y^0 & =  e^{ix^0}(s^0\cosh x^1 + is^1\sinh x^1),  &\quad
y^1 & =  e^{ix^0}(-is^0\sinh x^1 + s^1\cosh x^1), \\
y^2 & =  e^{ix^2}(s^2\cosh x^3 + is^3\sinh x^3),   &\quad
y^3 & =  e^{ix^2}(-is^2\sinh x^3 + s^3\cosh x^3). \tag 4.7
\endalignat
$$
Hamiltonian:
$$ \split
H_3= &\frac{1}{2}  (p_0^2 - p_1^2 + p_2^2 - p_3^2 +
 \frac{1}{((s^0)^2+(s^1)^2)^2}
[((s^0)^2-(s^1)^2)(k_0^2-k_1^2)+ 4s^0s^1k_0k_1] \\
& \quad + \frac{1}{((s^2)^2+(s^3)^2)^2}
[((s^2)^2-(s^3)^2)(k_2^2-k_3^2)+ 4s^2s^3k_2k_3] ).
\endsplit \tag 4.8
$$
The potential $V_3(s)$ is singular along two hyperbolas,
$s^0=s^1=0$  and $s^2=s^3=0$.
\bigskip
\noindent
{\bf 4.- The Orthogonally Decomposable MASA $M_4$; Metric
$K_2$.}
\medskip \noindent
Coordinates:
$$
y^0  =  e^{ix^0}s^0, \quad
y^1 =   e^{ix^1}s^1, \quad
y^2  =  e^{ix^2}(s^2 + ix^3s^3),\quad
y^3  =  e^{ix^2}s^3. \tag 4.9
$$
Hamiltonian:
$$
H_4= \frac 12 \left (p_0^2 - p_1^2 +2p_2p_3 +
 \frac{k_0^2}{(s^0)^2}-\frac{k_1^2}{(s^1)^2}+
2\frac{k_3(k_2s^3-k_3s^2)}{(s^3)^3} \right ). \tag 4.10
$$
The singularity surfaces $s^0=0$ and $s^1=0$ are
2--dimensional  hyperboloids; $s^3=0$ is a 2--dimensional
hyperbolic cylinder
$(s^0)^2-(s^1)^2=1, \,  s^3=0, \, s^2 \in \Bbb R$.
\bigskip
%\vfill\eject
\noindent
{\bf 5.- The Orthogonally Decomposable MASA $M_5$; Metric
$K_2$.}
\medskip\noindent
Coordinates:
$$
\alignat 2
y^0 & =  e^{ix^0}(s^0\cosh x^1 + is^1\sinh x^1),  &\quad
y^1 & =  e^{ix^0}(-is^0\sinh x^1 + s^1\cosh x^1), \\
y^2 & =  e^{ix^2}(s^2 + ix^3s^3),   &\quad
y^3 & =  e^{ix^2}s^3. \tag 4.11
\endalignat
$$
Hamiltonian:
$$
H_5= \frac 12 \left (p_0^2 - p_1^2 + 2p_2p_3 +
 \frac{((s^0)^2-(s^1)^2)(k_0^2-k_1^2)+4s^0s^1k_0k_1}{((
s^0)^2+(s^1)^2)^2} +
  2\frac{k_3(k_2s^3-k_3s^2)}{(s^3)^3}
\right ). \tag 4.12
$$

The potential $V_5(s)$ is singular on a one--dimensional
hyperbola
$s^0=s^1=0$ and on a 2--dimensional hyperbolic cylinder
$s^3=0, \, (s^0)^2-(s^1)^2=1,  \, s^2 \in \Bbb R$.
\bigskip
\noindent
{\bf 6.-  The OD MASA $M_6$; Metric $K_3$.}
\medskip\noindent
Coordinates:
$$
\alignat 2
y^0 & =  e^{ix^0}(s^0 + ix^1s^1),  &\quad
y^1 & =  e^{ix^0}s^1, \\
y^2 & =  e^{ix^2}(s^2 + ix^3s^3), & \quad
y^3 & =  e^{ix^2}s^3. \tag 4.13
\endalignat
$$
Hamiltonian:
$$
H_6= \left (p_0p_1 + p_2p_3 +
 \frac{k_1(k_0s^1-k_1s^0)}{(s^1)^3} +
  \frac{k_3(k_2s^3-k_3s^2)}{(s^3)^3}
\right ). \tag 4.14
$$
The potential is singular along the two hyperbolic
cylinders
$s^1=0$ and $s^3=0$.
\vfill\eject
%\bigskip
\noindent
{\bf 7.- The OD MASAs $M_7$ and $M_8$; Metric $K_{4,
\epsilon}$.}
\medskip\noindent
Coordinates:
$$
\alignat 2
y^0 & =  e^{ix^0}(s^0 + ix^1s^1 + (ix^2-\frac 12 (x^1)^2)
s^2),  &\quad y^1 & =  e^{ix^0}(s^1 +ix^1s^2), \\
y^2 & =  e^{ix^0}s^2 ,   &\quad
y^3 & =  e^{ix^0}s^3. \tag 4.15
\endalignat
$$
Hamiltonian:
$$
H_{7,8}= \frac \epsilon2 \left (2p_0p_2 + p_1^2-p_3^2 +
 \frac{k_1^2+2k_0k_2}{(s^2)^2} -
  \frac{4k_1k_2s^1}{(s^2)^3}+k_2^2\frac{(s^1)^2-2s^0s^2}{(
s^2)^4}-
\frac{k_3^2}{(s^3)^2} \right ). \tag 4.16
$$
The potential is singular along the hyperboloid $s^3=0$ and
the hyperbolic cylinder $s^2=0$.
%\vfill\eject
\bigskip
\noindent
{\bf 8.- The OID and D MASA $M_9$; metric $K_5$.}
\medskip\noindent
Coordinates:
$$\align
y^0 & =  e^{ix^0}([s^0+i(s^2x^2 + s^3x^3)]\cosh x^1 +
i[s^1+i(s^2x^3 -s^3x^2)]\sinh x^1),  \\
y^1 & =  e^{ix^0}(-i[s^0+i(s^2x^2 + s^3x^3)]\sinh x^1 +
i[s^1+i(s^2x^3 -s^3x^2)]\cosh x^1), \\
y^2 & =  e^{ix^0}(s^2\cosh x^1 - is^3\sinh x^1),  \tag 4.17
\\ y^3 & =  e^{ix^0}(is^2\sinh x^1 + s^3\cosh x^1).
\endalign
$$
Hamiltonian:
$$ \split
H_9= & (p_0p_2 + p_1p_3) +
 \frac{(k_0k_2+k_1k_3)((s^2)^2-(s^3)^2) +
  2(k_0k_3-k_1k_2)s^2s^3}{((s^2)^2+(s^3)^2)^2} \\ & +
\frac{1}{((s^2)^2+(s^3)^2)^3}\{(k_2^2-k_3^2)[s^1s^3(3(s^2)^2-
(s^3)^2)- s^0s^2((s^2)^2-3(s^3)^2)] \\
& -2k_2k_3[s^1s^2((s^2)^2-3(s^3)^2)+
s^0s^3(3(s^2)^2-(s^3)^2)]\} .  \endsplit \tag 4.18
$$
This Hamiltonian is actually {\sl nonsingular}, since the
point $s^2=s^3=0$  does not lie on the hyperboloid
$2(s^0s^2+s^1s^3)=1$.
\bigskip
\noindent
{\bf 9.- The MANS $M_{10}$; metric $K_6$.}
\medskip\noindent
Coordinates:
$$
\alignat 2
y^0 & =  e^{ix^0}(s^0+is^1x^1 +is^2x^2 +
(ix^3-x^1x^2)s^3),  &\quad
y^1 & =  e^{ix^0}(s^1+ix^2s^3), \\
y^2 & =  e^{ix^0}(s^2 + ix^1s^3),   &\quad
y^3 & =  e^{ix^0}s^3.\tag 4.19
\endalignat
$$
Hamiltonian:
$$
H_{10}= (p_0p_3 + p_1p_2) +
 \frac{k_0k_3+k_1k_2}{(s^3)^2} -
  \frac{2k_3}{(s^3)^3}(k_1s^2+k_2s^1) +
\frac{k_3^2}{(s^3)^4}(-s^0s^3+3s^1s^2). \tag 4.20
$$
The singularity surface of the potential is the hyperbolic
cylinder
$s^3=0, \, 2s^1s^2=1, \, s^0 \in \Bbb R$.
\bigskip
\noindent
{\bf 10.- The MANS $M_{11}$; Metric $K_6$.}
\medskip \noindent
Coordinates:
$$
\align
y^0 & =  e^{ix^0}(s^0+is^1x^1 +(ix^2- \frac{(x^1)^2}{2})s^2
+  [i(x^3- \frac{(x^1)^2}{6})- x^1x^2]s^3),  \\
y^1 & =  e^{ix^0}(s^1+ix^1s^2 +(ix^2- \frac{(x^1)^2}{2})s^3),
\\ y^2 & =  e^{ix^0}(s^2 + ix^1s^3),  \tag 4.21 \\
y^3 & =  e^{ix^0}s^3.
\endalign
$$
Hamiltonian:
$$ \split
H_{11}= & (p_0p_3 + p_1p_2) +
 \frac{k_0k_3+k_1k_2}{(s^3)^2} -
  \frac{(k_3)^2s^0+k_2^2s^2+2k_3(k_1s^2+k_2s^1)}{(s^3)^3}
\\ &\quad +
\frac{3k_3s^2(k_3s^1+k_2s^2)}{(s^3)^4}-
\frac{2k_3^2(s^2)^3}{(s^3)^5}.
\endsplit \tag 4.22
$$
The singularity surface of the potential is the hyperbolic
cylinder
$s^3=0, \, 2s^1s^2=1, \, s^0 \in \Bbb R$.

%%%%%%%%%%%%%%%%--CAPITULO V--%%%%%%%%%%%%%%%%%%%%%%%%%
\bigskip\bigskip
\centerline {\bf 5.- SUPERINTEGRABILITY} \smallskip
\centerline{\bf OF THE REDUCED HAMILTONIAN SYSTEMS.}
\bigskip \noindent
{\bf 5.1.- Complete Sets of Commuting Operators.}
\medskip
In order to show that the Hamiltonian $H$ of eq. (1.2) is
completely integrable, we need to present two integrals of
motion, say $Q_1$ and $Q_2$,  that are in involution and are
well defined on the phase space $\{s,p\}$.  Indeed, in view
of the constraint (1.1), we have just three degrees of
freedom, hence the set
$$
\{ H, \, Q_1,\, \, Q_2\} \tag 5.1
$$
guarantees completely integrability. Maximal
superintegrability, as defined in the introduction, requires
the existence  of $2n-1=5 $ functionally independent
integrals of motion, pairwise in  involution. We shall show
that each of the Hamiltonians $H_1, \dots, H_{11}$
constructed in Section 4 is maximally superintegrable.
\smallskip
To do this we return to the free Hamiltonian (1.5) on the
complex hyperboloid (1.6). The underlying configuration space
is seven--dimensional. The Hamiltonian system is hence
completely integrable, if there exist seven functionally
independent integrals of motion in involution. Five of them
we already know, namely the Hamiltonian $H$ proportional to
the $SU(2,2)$  second order Casimir operator and four basis
elements of the considered  MASA of $u(2,2)$. Thus, to the
MASA $\{ Y_1,\, Y_2,\, Y_3\}$ of $su(2,2)$ we add  the
diagonal element
$$
Y_0=iI_4, \tag 5.2
$$
represented by the vector field
$$
\hat {Y}_0 = -i(y^0\partial y^0
+ y^1\partial y^1 + y^2\partial y^2 + y^3\partial y^3).
\tag 5.3
$$
Functions $f(y)$ that project properly onto $\ph$ satisfy
$$
\hat {Y}_0 f(y)= 0.  \tag 5.4
$$

We hence need to find two more integrals of motion $T_1$
and $T_2$ to  form the complete set
$$
\{ H, \, Y_0, \, Y_1,\, Y_2, \, Y_3,\, T_1,\, T_2\}. \tag
5.5
$$
Since we wish to solve the Hamilton--Jacobi, or
Schr\"odinger equation  for the reduced Hamiltonian system
by separation of variables, we shall  require that $T_1$ and
$T_2$ be second order operators in the enveloping  algebra
[20--23] of $su(2,2)$:
$$
T_\alpha=\sum_{j=1}^{15}\sum_{i \leq j}a^\alpha_{ij}
(X^iX^j+X^jX^i), \quad \alpha=1,2,\,\, a^\alpha_{ij} \in
\Bbb R. \tag 5.6
$$
The first order operator $Y_0$ is fixed once and for all.
The first order operators $\{ Y_1,\, Y_2, \, Y_3 \}$ form a
basis for the MASA
$M_a,\,\,  a=1,2, \dots, 11$ and are given for each MASA in
column 3 of  Table 1. Constructing the operators $T_\alpha$,
satisfying
$$
[Y_i, \, T_\alpha]=0, \quad i=1,2,3, \,\, \alpha=1,2,   \tag
5.7
$$
is then a problem of linear algebra ($[Y_0, \, T_\alpha]=0$
is satisfied automatically).
\smallskip
The construction of the operators $T_\alpha$ can also be
viewed as a  problem of group representation theory. Indeed,
what is needed, is to find the invariants of the action of
the abelian group $G_a= \exp M_a$ on the  enveloping algebra
of $su(2,2)$. They can be obtained by considering the
coadjoint action of $G_a$ on the dual $su(2,2)^*$ of the Lie
algebra
$su(2,2)$:
$$
V^\prime=e^Y V e^{-Y}, \quad Y \in M_a, \, V \in su(2,2)^*.
\tag 5.8
$$
This was done for all Cartan subalgebras of $su(p,q)$ in
Ref.1.
\smallskip
Here we skip all details and simply present, in Appendix 1,
the sets of second order invariants for each of the MASAs
$M_a$. To each set given  in Appendix 1 we must add the six
operators $Y_iY_k, \, (i, k=1, 2, 3)$ in  the enveloping
algebra of the considered MASA, in order to obtain  a
complete set of second order invariants. In each case, one
linear  combination  of the invariants is equal to the second
order Casimir operator of $su(2,2)$.  The notation in
Appendix 1 is adapted to the  metric used for the MASA in
Table 1. Thus,  say for $M_6$, the metric is
$K_3$ and the elements
$X_i, \,  (i=1,\dots, 15)$, are to be read off from eq.
(2.3c).
\smallskip
The considered MASAs are all three--dimensional. The
corresponding abelian subgroups act on the 15 dimensional
space $su(2,2)^*$. We hence expect  to have 12 functionally
independent invariants in each case, though  they do not,
{\sl a priori}, have to be second order polynomials.  In all
cases, except $M_{10}$, we do get 12 second order
invariants.  For $M_{10}$ we get 13 of them; however a
polynomial relation between them  exists (a ``syzygy"),
making it possible to express a power of one of them  as a
polynomial in the others.
\smallskip
Obviously the second order operators $T_\alpha$, obtained
for a given MASA,  do not all commute amongst each other.
Their commutation relations are  given in Appendix 2. The
letters $A,\, \dots,\, D$ denote certain third  order
polynomials in the enveloping algebra of $su(2,2)$. Their
actual  form is immaterial, except for the fact that they
are always linearly  independent.
\smallskip
{}From Appendix 2 we see that it is possible to choose
commuting pairs
$\{ T_1, \, T_2\}$ in many different ways and thus to obtain
different  complete sets of commuting operators (5.5). The
operators $T_1$ and
$T_2$ will restrict to the operators $Q_1$ and $Q_2$ of eq.
(5.1)  upon reduction to the $O(2,2)$ hyperboloid. Maximal
superintegrability  is assured by having (at least) two
different pairs $\{ T_1, \, T_2\}$  and $\{ T_3, \, T_4\}$
for a given MASA, such that the four are linearly
independent and satisfy
$$
[T_1, \, T_2]=0, \qquad [T_3, \, T_4]=0.  \tag 5.9
$$

A systematic search for commuting pairs of operators $[T_a,
\, T_b]=0$ is  related to the problem of separating variables
in Hamilton--Jacobi and  Laplace--Beltrami equations on the
$O(2,2)$ hyperboloid $\sim \ho$. This  problem was solved by
Kalnins and Miller [24] (see also Ref.[20, 25--27]).  They
obtained 74 families of separable coordinates (3 of them
nonorthogonal) and classified the corresponding second order
operators. The number 74 is actually somewhat ambiguous,
since sometimes  several systems complement  each other to
cover the entire $O(2,2)$  hyperboloid and some systems are
related to others by an outer  automorphism of $O(2,2).$
\smallskip
In order to classify commuting pairs of operators $\{ Q_1, \,
Q_2\}$  obtained by projecting the commuting pairs $\{T_1, \,
T_2\}$, we shall  need a classification of $O(2,2)$ pairs
of operators and separable  coordinates [20, 24].
\smallskip
We recall that an ``ignorable variable" is one that does
not figure in the metric tensor $g_{ik}(s)$ written in the
corresponding separable coordinate  system.
\smallskip
For $O(2,2)$ we shall now use the metric $g=K_1$. A basis
for the $O(2,2)$  algebra is given by
$$\alignat 2
K_{01}&=s^0\partial_{s^1}+s^1\partial_{s^0},& \qquad
K_{23}&=s^2\partial_{s^3}+s^3\partial_{s^2}, \\
K_{03}&=s^0\partial_{s^3}+s^3\partial_{s^0}, &\qquad
L_{02}&=s^0\partial_{s^2}-s^2\partial_{s^0}, \tag 5.10 \\
K_{12}&=s^1\partial_{s^2}+s^2\partial_{s^1},&\qquad
L_{13}&=s^1\partial_{s^3}-s^3\partial_{s^1}.
\endalignat
$$
The types of coordinates and commuting operators that occur
for
$O(2,2)/O(2,1)$ are the following:
\bigskip \noindent
I.- \underbar{\sl Two ignorable variables.}
\medskip
The operators $Q_1$ and $Q_2$ are squares of elements of a
MASA of $o(2,2)$.  Six inequivalent MASAs exist, yielding the
pairs
$$\align
&1.- \, (L^2_{02},\,  L^2_{13}), \\
&2.- \, (K^2_{01},\, K^2_{23}), \\
&3.- \, ((L_{02}+K_{03})^2,\, (L_{13}+K_{12})^2), \tag 5.11
\\ &4.- \, ((L_{02}-L_{13})^2,\, (K_{12}+K_{03})^2),\\
&5.- \, ((L_{02}-L_{13})^2,\, (L_{02}+L_{13}+K_{12}+K_{03}
)^2),\\ &6.- \,((K_{12}-K_{03})^2,\,
(L_{02}+L_{13}+K_{12}+K_{03})^2).
\endalign $$
The first three types correspond to the group reductions
$O(2,2) \supset O(2) \times O(2), \, \, O(2,2) \supset O(1,1)
\times  O(1,1)$  and $O(2,2) \supset E(1) \times E(1)$,
respectively. The separable  coordinates are orthogonal
($E(1)$ is the one--dimensional Euclidean group,  i.e.,
translations generated e.g. by $(L_{02}+K_{03}$)). The last
three  types correspond to the reductions $O(2,2) \supset
O(2) \times O(1,1),
\, \, O(2,2) \supset O(2) \times T(1)$ and $O(2,2) \supset
O(1,1)
\times T(1)$, respectively, the coordinates are
nonorthogonal.
\bigskip \noindent
 II.- \underbar{\sl One ignorable variable.}
\medskip
A). {\sl Subgroup type}.
\smallskip
The operator $Q_1$ is the square of an element of the Lie
algebra $o(2,2)$,
$Q_2$ is the Casimir operator of a subalgebra of $o(2,2)$,
i.e. either
$e(1,1)$ (the Poincar\'e algebra in one space dimension),
or $o(2,1)$.

The corresponding pairs are:
$$\alignat 2
&7.- \, (K_{01}^2,\, (L_{02}+K_{03})^2-(K_{12}+L_{13})^2),
&\quad  &O(2,2) \supset E(1,1) \supset  O(1,1),  \\
&8.- \, (L_{02}^2,\, K_{01}^2+K_{12}^2-L_{02}^2), &\quad
&O(2,2) \supset O(2,1)  \supset O(2),\tag 5.12\\
&9.- \,(K_{01}^2,\, K_{01}^2+K_{12}^2-L_{02}^2), &\quad
&O(2,2) \supset O(2,1)  \supset O(1,1),\\
&10.- \,((K_{01}+L_{02})^2,\, K_{01}^2+K_{12}^2-L_{02}^2),
&\quad &O(2,2) \supset O(2,1)  \supset E(1).
\endalignat
$$
We note here that the algebra $o(2,2)$ allows an outer
automorphism  realized, e.g., by the permutation of indices
$$
0 \leftrightarrow 1, \qquad  2\leftrightarrow 3  \tag 5.13
$$
in the coordinates $s^\mu$ and vector fields (5.10). This
permutation may  provide new coordinate systems and new
commuting pairs of operators.  This is the case of
eq.(5.12),  when the $O(2,1)$ subgroup is replaced by
$O(1,2)$. In the case of relation (5.12) the permutation
(5.13) provides an  equivalent system (i.e. the new one can
be rotated back into the old one).  We shall not analyze this
question any further. Below it is to be  understood  that the
permutation (5.13) should be applied whenever it yields new
systems  (they are always very similar to those listed).
\medskip
B). {\sl Generic Type}.
\smallskip
The operator $Q_1$ is the square of an element of $o(2,2)$,
whereas $Q_2$  is generic, i.e., not a Casimir operator of
any Lie subgroup.
\medskip
B$_1$). {\sl The $O(2)$ element $L_{02}$} yields
$$ \split
11.- &\,(L_{02}^2,\, K_{01}^2+K_{12}^2 +a(K_{03}^2+K_{23}^2)
+ bL_{13}^2+c(\{ K_{01}, \, K_{23}\}-\{K_{12}, \, K_{03} \})
+ \\ &\qquad d(\{ K_{01}, \, K_{03}\}+\{K_{12}, \, K_{23}
\})).
\endsplit \tag 5.14
$$
The curly brackets \{ \, , \,  \} denote anticommutators.
Separable  coordinates  [24] are obtained for
$$
b=c=d=0, \qquad \text{or}\quad a=c=d=0. \tag 5.15
$$
The permutation (5.13) provides further systems, but we
shall not discuss them here (nor below).
\smallskip
B$_2$). {\sl The $o(1,1)$ element $K_{01}$} yields
$$\split
12.- &\,(K_{01}^2,\, a(K_{12}^2-L_{02}^2) +
bK_{23}^2+c(K_{03}^2-L_{13}^2)+ \\
&\qquad d(\{ L_{13}, \, K_{12}\}-\{L_{02}, \, K_{03} \}) +
e(( L_{02}-K_{03})^2-(L_{13}-K_{12})^2)).
\endsplit \tag 5.16
$$
Different separable coodinates are obtained for
$$
\alignat 3
a&=1, &\qquad c&=d=e=0, \quad &  \tag 5.17a \\
a&=1, &\qquad b&=d=e=0, \quad &   \tag 5.17b \\
d&=1, &\qquad a&=c, \quad &b=e=0,  \tag 5.17c \\
b&=1, &\qquad e&=1, \quad &a=c=d=0.  \tag 5.17d
\endalignat
$$
\medskip
B$_3$). {\sl The $e(1)$ element $L_{02}+K_{12}$} yields:
$$\split
13.- &\,((L_{02}+K_{12})^2,\, a(K_{12}^2+K_{01}^2-L_{02}^2)
+ b(L_{13}+K_{03})^2+ \\
&\qquad c(\{ K_{23}, \, L_{02}+K_{12}\}+\{ K_{01}, \, L_{13}
+K_{03}\})).
\endsplit \tag 5.18
$$
Different separable coordinates are obtained for:
$$\alignat 2
a&=1, \quad b=\pm 1, &\qquad c&=0,  \tag 5.19a \\
a&=b=0, &\qquad c&=1.    \tag 5.19b
\endalignat
$$
Other pairs of commuting operators satisfying $Q_1=X^2, \,
\, X\in o(2,2)$, exist,  but they do not correspond to
separable coordinates on the $O(2,2)$  hyperboloid.
\bigskip
\noindent
III. \underbar{\sl No ignorable variables}.
\medskip
A). {\sl Subgroup type.}
\smallskip
The operator $Q_1$ is a Casimir operator of a subalgebra of
$o(2,2)$, i.e.
$e(1,1)$ or $o(2,1)$. The operator $Q_2$ lies in the
enveloping algebra  of the corresponding subalgebra and is
not the square of an element of
$o(2,2)$.
\medskip
A$_1$). $$e(1,1)=(K_{01}, \, L_{02}+K_{03}, \, L_{13}+K_{12}
).$$ Notice that the $o(2,2)$ element $K_{23}$ acts like a
dilation on the translations $L_{02}+K_{03}\equiv P_0$ and
$L_{13}+K_{12} \equiv P_1$. We have
$$
Q_1=(L_{02}+K_{03})^2- (L_{13}+K_{12})^2 \equiv P_0^2-P_1^2,  \tag 5.20
$$
and $Q_2$ runs through 8 possibilities:
$$
K_{01}^2  \pm(P_0^2+P_1^2), \, \, \, K_{01}^2+2P_0P_1, \, \,
\,  K_{01}^2 \pm(P_0+P_1)^2,  \tag 5.21
$$
$$
\lbrace K_{01}, \, P_0 \rbrace , \, \, \, \lbrace K_{01}, \,
P_1\rbrace ,
\,\, \, \lbrace K_{01}, \,\,  P_0-P_1\rbrace  +(P_0+P_1)^2.
$$
The ninth possibility $Q_2=\lbrace K_{01}, \, P_0-P_1\rbrace
$ does not  correspond to separable coordinates [24--27].
\medskip
A$_2$).
$$
o(2,1)=(K_{01}, \,K_{12}, \,L_{02}). \tag 5.22
$$

We have
$$
Q_1=K_{01}^2+K_{12}^2-L_{02}^2, \tag 5.23
$$
and $Q_2$ runs through the following cases [22]:
$$
K_{01}^2+aK_{12}^2, \, \, \lbrace L_{02}, \, K_{01}\rbrace
+a (L_{02}^2-K_{01}^2),  \, \, K_{01}^2 \pm (K_{12}+L_{02}
)^2, \, \,
\lbrace K_{01}, \, K_{12}+L_{02}\rbrace .  \tag 5.24
$$
\medskip
B). {\sl Generic Type.}
\smallskip
The operators $Q_1$ and $Q_2$ are neither perfect squares,
nor Casimir  operators of subalgebras of $o(2,2)$.
\smallskip
Essentially 9 different classes of such commuting pairs
exist. For details we refer to Kalnins and Miller [24].
\bigskip \noindent
{\bf 5.2.- Restriction of the su(2,2) Lie Algebra and of
the Integrals of  Motion to the O(2,2) Hyperboloid.}
\medskip
In Section 3 we presented general formulas for reducing  the
elements of the Lie algebra of $su(2,2)$ to the real $O(2,2)$
hyperboloid, namely eq.(3.9)  and (3.10). In Section 4 we
introduced the real coordinates $(s,x)$  for each MASA of
$su(2,2)$ and used them to reduce the $SU(2,2)$ free
Hamiltonian to 11 different $O(2,2)$ Hamiltonians with
nontrivial  potentials.
\medskip
Let us now perform a similar reduction for the generators
$\hat X_i$ of
$su(2,2)$ and for the integrals of motion $T_a$ of Appendix
1. Generally  speaking the integral $T_a$ after reduction
will have the form
$$
\widetilde T_a=\widetilde T_a(\text{kin}) +\widetilde T_a(
\text{pot}),
\tag 5.25
$$
where the ``kinetic" part comes from products of terms like
(3.9), the ``potential" part from terms like (3.10). We will
be mainly interested in
$\widetilde T_a(\text{kin})$. It lies in the enveloping
algebra of
$o(2,2)$ and determines the separable coordinate systems.
\medskip
Note that if two operators $(T_a, \, T_b)$ in the enveloping
algebra of
$su(2,2)$  commute, then so do the corresponding reduced
operators
$(\widetilde T_a, \, \widetilde T_b)$. In this case, if  the
``kinetic"  parts $(\widetilde T_a(\text{kin}), \,
\widetilde T_b(
\text{kin}))$ correspond to a certain coordinate system  for
which the free $O(2,2)$ Hamiltonian allows the separation of
variables, then the reduced  Hamiltonian with the induced
potential also allows the  separation of variables in the
same system.
\medskip
Let us now run through the individual MASAs. We shall use
the notations  (5.10) for the $o(2,2)$ basis elements, i.e.,
all the second order  integrals of motion are given in the
diagonal metric $g=K_1=
\text{diag}(1,-1,1,-1)$. We shall use the same notation for
the reduced invariants, as for the original ones, i.e., drop
the tildes of eq. (5.25).
\bigskip
\noindent
1.- {\bf The Compact Cartan Subalgebra $M_1$.}
\medskip
The invariants $T_1,\, \, \dots, \, T_6$ of Appendix 1
restricted to
$o(2,2)$  are
$$\alignat 2
T_{1} & = {K}_{01}^2+
(k_0\frac{s^1}{s^0}-k_1\frac{s^0}{s^1})^2,&\quad
T_{2} & = {L}_{02}^2+
(k_0\frac{s^2}{s^0}+k_2\frac{s^0}{s^2})^2,\\
T_{3} & = {K}_{03}^2+
(k_0\frac{s^3}{s^0}-k_3\frac{s^0}{s^3})^2,  &\quad
T_{4} & = {K}_{12}^2+
(k_1\frac{s^2}{s^1}-k_2\frac{s^1}{s^2})^2, \tag 5.26\\
T_{5} & =  {L}_{13}^2+
(k_1\frac{s^3}{s^1}+k_3\frac{s^1}{s^3})^2, &\quad
T_{6} &= {K}_{23}^2+
(k_2\frac{s^3}{s^2}-k_3\frac{s^2}{s^3})^2.
\endalignat
$$

The Hamiltonian (4.4) is recovered as
$$
H=\frac 12 (-T_1+T_2-T_3-T_4+T_5-T_6) +\sum k_\mu^2 -2
\sum_{\mu < \nu}k_\mu k_\nu.
$$

The following pairs of commuting operators and separable
$O(2,2)$  coordinates can be constructed:
\medskip
I. \underbar{\sl 2 ignorable variables.}
$$\align
&1.- \,\, (T_2, \, T_5), \\  &2.-\,\,(T_1, \, T_6) .  \tag
5.27
\endalign
$$

The coordinates are orthogonal and correspond to the
reductions
$O(2,2) \supset O(2) \times O(2)$ and $O(2,2) \supset
O(1,1) \times O(1,1)$, respectively.
\medskip
II. \underbar{\sl 1 ignorable variable.}
\smallskip
A). {\sl Subgroup type.}
$$ \alignat 2
&3.- \,\, (T_1+T_4-T_2, \, T_2), &\quad  &O(2,2) \supset
O(2,1)  \supset O(2),
\\
&4.-\,\,(T_1+T_4-T_2, \, T_1), &\quad  &O(2,2) \supset
O(2,1)  \supset O(1,1) .
\tag5.28
\endalignat
$$

B). {\sl Generic.}
$$\alignat 2
5.- \,\, (T_2,\,T_1+T_4+a(T_3+T_6)+bT_5), &\quad &a \neq 0
\tag 5.29 \\ 6.- \,\, (T_1,\,T_4-T_2+a(T_3-T_5)+bT_6) &
\quad &a \neq 0.
\endalignat
$$

III. \underbar{\sl No ignorable variables}.
\smallskip
 A). {\sl Subgroup type}
$$
7.- \,\, (T_1+T_4-T_2, \, T_1+aT_4), \quad a \neq 0.\tag
5.30
$$

B). {\sl Generic type}
$$
8.- \,\, (T_2+aT_1+bT_3, \, T_5+\frac{a-b}{b}T_1+
\frac{a-b}{(1+a)b}T_4),
\quad b \neq 0, \, a \neq -1. \tag 5.31
$$
%\vfill\eject\bigskip
2.- {\bf The Noncompact Cartan Subalgebra $M_2$.}
\medskip
The invariants restricted to $O(2,2)$ are
$$\align
T_{1} & =   K_{01}^2+(k_0\frac{s^1}{s^0} -k_1
\frac{s^0}{s^1})^2,  \tag 5.32\\  T_{2} & =
-K_{23}^2+\frac{1}{((s^2)^2+(s^3)^2)^2}[s^2(s^2k_2+s^3k_3)
+s^3(s^2k_3-s^3k_2)]^2 ,  \\
 T_{3} & =  L_{02}^2 - K_{03}^2+[k_0\frac{s^2}{s^0}+
\frac{s^0}{(s^2)^2+(s^3)^2} (s^2k_2 + s^3k_3)]^2
- [k_0\frac{s^3}{s^0}+\frac{s^0}{(s^2)^2+(s^3)^2} (s^2k_3  -
s^3k_2)]^2,
\\ T_{4} & =
K_{12}^2-L_{13}^2+[k_1\frac{s^2}{s^1}-\frac{s^1}{(s^2)^2+
(s^3)^2}   (k_2s^2 + k_3s^3)]^2
-[k_1\frac{s^3}{s^1}-\frac{s^1}{(s^2)^2+(s^3)^2} (s^2k_3
-s^3k_2)]^2,  \\ T_{5} & =  -\{ L_{02}, \, K_{03} \}
-2[k_0\frac{s^2}{s^0} +
\frac{s^0}{(s^2)^2+(s^3)^2} (s^2k_2 + s^3k_3)]
[k_0\frac{s^3}{s^0} +\frac{s^0}{(s^2)^2+(s^3)^2} (s^2k_3 -
s^3k_2)],  \\ T_{6} & =  -\lbrace K_{12}, \, L_{13} \rbrace
+2[k_1\frac{s^2}{s^1}-
\frac{s^1}{((s^2)^2+(s^3)^2)}(s^2k_2+s^3k_3)][k_1
\frac{s^3}{s^1}-
\frac{s^1}{((s^2)^2+(s^3)^2)}(s^2k_3-s^3k_2)].
\endalign
$$

The Hamiltonian, up to additive and multiplicative
constants, is given by
$-T_1+T_2+T_3-T_4$.
\smallskip
The commuting pairs of second order integrals of motion,
and the  coordinates are as follows.
\medskip
I. \underbar{\sl 2 ignorable variables}
$$
1.-\,\,(T_1,\,T_2), \qquad O(1,1) \times O(1,1). \tag 5.33
$$

II. \underbar{\sl 1 ignorable variable}
\smallskip
A). {\sl Subgroup type}
$$
2.-\,\,(T_2+T_3,\,T_2), \qquad O(2,2) \supset O(2,1) \supset
O(1,1).
\tag 5.34
$$

B). {\sl Generic.}
$$\align
&3.-\,\,(T_1,\, a(T_3-T_4) + b(T_5-T_6)), \\
&4.-\,\,(T_2,\,aT_3+bT_4),\quad ab \neq 0.
\tag 5.35 \endalign
$$

III. \underbar{\sl No ignorable variables}.
\smallskip
 A). {\sl Subgroup type.}
$$
5.-\,\,(T_2+T_3,\,T_5+aT_2). \tag 5.36
$$

B). {\sl Generic.}
$$ \split
6.-\,\,(T_1+ &aT_3+bT_5,\,T_2+\frac{b^2}{a(a+1)+b^2}T_3-
\frac{(a+1)b}{a(a+1)+b^2}T_5 \\
&+\frac{b}{a(a+1)+b^2}T_6), \qquad
a^2+b^2+a \neq 0, \endsplit
$$
$$
7.-\,\,(T_1-T_3,\,T_2+aT_6), \quad a \neq 0, \tag 5.37
$$
$$
8.-\,\,(T_1+aT_3+bT_5+cT_6, \, (b+c+ac)T_3+(bc-a-1)T_5+(1+
c^2)T_6),
$$
with $a+a^2+b^2+bc = 0.$
\bigskip
{\bf 3.- The Noncompact Cartan Subalgebra $M_3$.}
\medskip
{}From here on we shall spell out the ``kinetic" parts of the
invariants  only; the remaining ``potential" parts are easy
to calculate (and are available from the authors upon
request).
$$\alignat 2
T_1&=-K^2_{01} +f_1(s),&\quad
T_2&=-K^2_{23}+f_2(s), \tag 5.38\\
T_3&=L^2_{02}-K^2_{03}-K^2_{12}+L^2_{13}+f_3(s),&\quad
T_4&=-\{ L_{02},\,K_{03} \} +\{ K_{12},\,L_{13} \}+f_4(s),\\
T_5&=-\{ L_{02},\,K_{12}\}+\{ K_{03},\,L_{13} \}+f_5(s),
&\quad  T_6&=\{ L_{02},\,L_{13}\}+\{ K_{03},\,K_{12}
\}+f_6(s).  \endalignat
$$

The Hamiltonian satisfies $H \thicksim T_1+T_2+T_3.$
\smallskip
Commuting pairs and coordinates:
\medskip
I. \underbar{\sl 2 ignorable variables}.
$$
1.-\,\,(T_1,\,T_2),\qquad O(1,1) \times O(1,1). \tag 5.39
$$

II. \underbar{\sl 1 ignorable variable}.
\smallskip
 A). {\sl Subgroup type}: None
\smallskip
B). {\sl Generic}:
$$
2.-\,\,(T_1,\,T_4+aT_2). \tag 5.40
$$

III. \underbar{\sl No ignorable variables}.
\smallskip
A). {\sl Subgroup type}: None.
\smallskip
B). {\sl Generic}:
$$ 3.-\,\,Q_1= T_1+aT_4+bT_5+cT_6, \quad Q_2= T_2-T_1+
\frac{bc-a}{c^2}T_4 -\frac{a}{c}T_5,$$
with $c \neq 0$ and $c(a^2+b^2+c^2)-ab=0$,
$$ 4.-\,\,Q_1= T_1+a^2T_2+aT_6, \quad Q_2= T_4+aT_5,
\quad a
\neq 0.
\tag 5.41
$$
\bigskip
{\bf 4. The OD MASA $M_4$.}
\medskip
The reduced integrals of motion are
$$\alignat 2
T_1&=K^2_{01} +f_1(s),&\quad
T_2&=\frac12(L_{02}+K_{03})^2+f_2(s), \\
T_3&=\frac12(L_{13}+K_{12})^2+f_3(s),&\quad
T_4&=K_{23}^2 +f_4(s),\tag 5.42 \\
T_5&=L_{02}^2-K_{03}^2+f_5(s), &\quad
T_6&=-L_{13}^2+K_{12}^2+f_6(s) .
\endalignat $$
We have $H \thicksim T_1+T_4-T_5+T_6.$
\smallskip
The commuting pairs are:
\smallskip
I. \underbar{\sl 2 ignorable variables}.
$$\alignat 2
&1.-\,\,(T_1,\,T_4), &\qquad &O(1,1) \times O(1,1), \\
&2.-\,\,(T_2,\,T_3), &\qquad &E(1) \times E(1). \tag 5.43
\endalignat
$$
\vfill\eject
II. \underbar{\sl 1 ignorable variable}.
\smallskip
A). {\sl Subgroup type.}
$$
\alignat 2
&3.-\,\,(T_5-T_4,\,T_4), &\qquad &O(2,2) \supset O(2,1)
\supset O(1,1), \\  &4.-\,\,(T_5-T_4,\,T_2), &\qquad &O(2,2)
\supset O(2,1) \supset E(1),
\tag 5.44 \\
&5.-\,\,(T_2-T_3,\,T_1), &\qquad &O(2,2) \supset E(1,1)
\supset O(1,1).
\endalignat
$$

B). {\sl Generic}.
$$ \alignat 2
&6.-\,\,(T_1, \,T_4+a(T_2-T_3)), &\qquad &a \neq 0,\\
&7.-\,\,(T_4, \,T_1+aT_5), &\qquad &a \neq 0,\tag 5.45\\
&8.-\,\,(T_2, \,T_4-T_5+aT_3), &\qquad &a \neq 0,
\endalignat
$$

III. \underbar{\sl No ignorable variables}.
\smallskip
A). {\sl Subgroup type}.
$$\alignat 2
&9.-\,\,(T_2-T_3, \,T_1+aT_2), &\qquad &a \neq 0,\\
&10.-\,\,(T_5-T_4, \,T_4+aT_2), &\qquad &a \neq 0.  \tag 5.46
\endalignat
$$

B). {\sl Generic}.
$$
11.-\,\,(T_1+aT_3+b(T_4-T_5), \,a(b-1)T_2+aT_3+bT_4),
\quad a(b-1) \neq 0.  \tag 5.47
$$
\bigskip
{\bf 5. The OD MASA $M_5$}.
\medskip
The integrals of motion now are:
$$
\alignat 2
T_1&=-K^2_{01} +f_1(s),&\quad
T_2&=K_{23}^2+f_2(s), \tag 5.48 \\
T_3&=\frac12[(L_{02}-K_{03})^2- (L_{13}-K_{12})^2]+f_3(s)
,&\quad  T_4&=\frac12 \{ L_{02}-K_{03}, \,-L_{13}+K_{12} \}
+f_4(s),  \\
T_5&=L_{02}^2-K_{03}^2+L^2_{13}-K^2_{12}+f_5(s), &\quad
T_6&=\{ L_{02}, \, K_{12} \}- \{ K_{03}, \, L_{13} \}+
f_6(s).
\endalignat
$$

We have $H \thicksim T_1-T_2+T_5.$
\medskip
The commuting pairs are:
\smallskip
I. \underbar{\sl 2 ignorable variables}.
$$ 1.-\,\, (T_1,\,T_2), \qquad O(1,1) \times O(1,1),
$$
$$ 2.-\,\, (T_3,\,T_4), \qquad E(1) \times E(1). \tag 5.49
$$

Note that while $T_3$ and $T_4$ are not directly squares  of
elements of a MASA, diagonalizing them is equivalent to
diagonalizing the commuting pair
$(L_{02}-K_{03},\, L_{13}-K_{12})$, a MASA of $o(2,2)$.
Similar situations with two ignorable variables occur below.
\medskip
II. \underbar{\sl 1 ignorable variable}.
\smallskip
A). {\sl Subgroup type}.
$$
3.-\,\,(T_3,\,T_1), \qquad O(2,2) \supset E(1,1) \supset
O(1,1). \tag 5.50
$$

B). {\sl Generic}.
$$
\alignat 2
&4.-\,\,(T_1, \,T_2+aT_3), &\qquad &a \neq 0,\\
&5.-\,\,(T_2, \,T_6+aT_1). \tag 5.51
\endalignat
$$

III. \underbar {\sl No ignorable variables}.
\smallskip
A). {\sl Subgroup type}.
$$
6.-\,\,(T_3, \,T_1+aT_4), \qquad a \neq 0. \tag 5.52
$$

B). {\sl Generic}.
$$ \alignat 2
&7.-\,\,(T_2+aT_3+bT_4, \,aT_1+b^2T_3-abT_4-bT_6), &
\qquad &a \neq 0,
\tag 5.53 \\
&8.-\,\,(T_2+aT_4, \,T_6-aT_3), &\qquad &a \neq 0.
\endalignat
$$

\bigskip
{\bf 6. The OD MASA $M_6$.}
\medskip
The integrals of motion:
$$
\alignat 2
T_1&=\frac 14 (L_{02}-K_{03}+K_{12}-L_{13})^2 +f_1(s),&
\quad  T_2&=K_{01}^2+f_2(s), \tag 5.54 \\
T_3&=K_{23}^2+f_3(s),&\quad
T_4&=\frac12 ((L_{02}+K_{12})^2-(K_{03}+L_{13})^2) +f_4(s),
\\   T_5&=\frac12
((L_{02}-K_{03})^2-(K_{12}-L_{13})^2)+f_5(s), &\quad
T_6&=L_{02}^2+L_{13}^2- K_{03}^2-K_{12}^2+f_6(s). \\
\endalignat
$$

We have $H \thicksim T_6-T_2-T_3.$
\medskip
The commuting pairs are:
\smallskip
I. \underbar{\sl Two ignorable variables}.
$$ 1.-\,\,(T_2, \,T_3),\qquad O(2,2) \supset O(1,1) \times
O(1,1),
$$
$$ 2.-\,\,(T_1, \,T_4),\qquad O(2,2) \supset E(1) \times
E(1). \tag 5.55
$$
\vfill\eject
II. \underbar{\sl One ignorable variable}.
\smallskip
A). {\sl Subgroup type}.
$$
3.-\,\,(T_4,\,T_3), \qquad O(2,2) \supset O(2,1) \supset
O(1,1). \tag 5.56
$$

B). {\sl Generic}.
$$
\alignat 2
&4.- \,\,(T_2,\,T_3+aT_5), &\qquad &a \neq 0,  \\
&5. -\,\,(T_1,\,T_4+aT_5), &\qquad &a \neq 0. \tag 5.57
\endalignat
$$

III. \underbar {\sl No ignorable variables}.
\smallskip
A). {\sl Subgroup type}.
$$
6.-\,\,(T_4,\,T_3+aT_1). \tag 5.58
$$

B). {\sl Generic}.
$$
7.-\,\,(T_2+abT_1+aT_4, \,T_3+abT_1+bT_5), \qquad ab    \neq
0.
\tag 5.59
$$
\bigskip
{\bf 7. The OD MASAs $M_7$ and $M_8$.}
\medskip
The integrals of motion are:
$$
\alignat 2
T_1&=\frac12(L_{13}-K_{23})^2+f_1(s),&\quad
T_2&=\frac12(K_{01}+L_{02})^2+f_2(s), \\
T_3&=K_{03}^2+K_{23}^2-L_{13}^2+f_3(s),&\quad
T_4&=4 (K_{12}^2+K_{01}^2-L_{02}^2) +f_4(s), \tag 5.60 \\
T_5&=\frac{1}{\sqrt 2} \{ K_{03},\,-L_{13}+K_{23} \} +
f_5(s), &\quad   T_6&=\sqrt 2 \{ K_{12}, \, K_{01}+L_{02}
\}+f_6(s).
\endalignat
$$

We have $H \thicksim T_3+\frac 14T_4.$
\medskip
The commuting pairs are:
\smallskip
I. \underbar {\sl Two ignorable variables}.
$$
1.-\,\,(T_1, \,T_2),\qquad O(2,2) \supset E(1) \times E(1).
\tag 5.61
$$

II. \underbar {\sl One ignorable variable}.
\smallskip
A). {\sl Subgroup type}.
$$
2.-\,\,(T_4,\,T_2), \qquad O(2,2) \supset O(2,1) \supset
E(1). \tag 5.62
$$

B). {\sl Generic}.
$$\alignat 2
&3.-\,\,(T_1,\,2T_5+T_6+aT_2), & \tag 5.63 \\
&4.-\,\,(T_2, \,T_4+aT_1), &\qquad &a \neq 0.
\endalignat
$$

III. \underbar {\sl No ignorable variables}.
\smallskip
A). {\sl Subgroup type}.
$$ \align
&5.-\,\,(T_1-T_2, \,T_5+aT_1), \\
&6.-\,\,(T_4, \,T_6+aT_2), \quad a \neq 0. \tag 5.64
\endalign
$$

B). {\sl Generic}.
$$
7.-\,\,(T_1+aT_2+bT_6,\,b^2T_4+a(a+1)T_2+2bT_5+(a+1)bT_6),
\qquad b \neq 0. \tag 5.65
$$
\bigskip
{\bf 8. The OID \& D MASA $M_9$}
\medskip
The invariants are
$$
\align
T_1&=\frac14(L_{02}-K_{03}+K_{12}-L_{13})^2+f_1(s),\\
T_2&=(K_{01}+K_{23})^2-(L_{02}+L_{13})^2+f_2(s), \\
T_3&=(K_{01}-K_{23})^2+2(K_{03}^2+K_{12}^2)-(L_{02}-L_{13
})^2+f_3(s),\\  T_4&=\frac12 \{
K_{01}-K_{23},\,\,L_{02}-K_{03}+K_{12} -L_{13} \}  +f_4(s),
\tag 5.66 \\   T_5&=\frac12 \{
K_{03}+K_{12},\,L_{02}-K_{03}+K_{12}-L_{13} \}  +f_5(s),\\
T_6&= \{ K_{01}+K_{23}, \,\,L_{02}+L_{13} \}+f_6(s).
\endalign
$$

We have $H \thicksim T_2+T_3.$
\medskip
The commuting pairs are:
\smallskip I. \underbar {\sl Two ignorable variables.}
$$ 1.-\,\,(T_2,\,T_6), \quad O(2,2) \supset O(1,1) \times
O(2),
$$
$$ 2.-\,\,(T_1,\,T_4),\quad O(2,2) \supset O(1,1) \times
E(1). \tag 5.67
$$
 \smallskip II. \underbar {\sl One ignorable variable.}
\smallskip A). {\sl Subgroup type}: None.
\smallskip
B). {\sl Generic}
$$
3.-\,\,(T_1,\,T_4+aT_5), \qquad a\neq 0. \tag 5.68
$$

III. \underbar {\sl No ignorable variables}.
\smallskip
A). {\sl Subgroup type}: none.
\smallskip
B). {\sl Generic}
$$
4.-\,\,(-(a^2+b^2)T_1+T_2+aT_4+bT_5,\,-bT_4+aT_5+T_6), \quad
a^2+b^2 \neq 0.
\tag 5.69
$$

The first two pairs above correspond to nonorthogonal
coordinates on the
$O(2,2)$ hyperboloid [24]. The third does not correspond  to
any separable system.
\bigskip
{\bf 9. The MANS $M_{10}$}
\medskip
The situation is different in this case than for all other
MASAs of
$su(2,2)$. First of all, there are 7 second order integrals
in the enveloping algebra of
$su(2,2)$, rather them 6 as in all other cases. Secondly,
one of them,
$T_7$  in Appendix I, does not reduce to the $O(2,2)$
manifold as in eq. (5.25). Instead, it reduces to a first
order operator, once the momenta, conjugate to ignorable
variables are set equal to the constants $k_\mu$. Indeed,
after  the reduction the 7 invariants are:
$$
\align
T_1&=\frac14(K_{01}-L_{13}+L_{02}+K_{23})^2+f_1(s),\\
T_2&=\frac14(-K_{01}+L_{13}+L_{02}+K_{23})^2+f_2(s), \\
T_3&=\frac14((L_{02}+K_{23})^2-(K_{01}-L_{13})^2)+f_3(s),\\
T_4&=\frac12 \{ K_{03}+K_{12},\,\,K_{01}-L_{13}+L_{02}+
K_{23} \}  +f_4(s), \tag 5.70 \\
T_5&=\frac12 \{ K_{03}-K_{12},\,-K_{01}+L_{13}+L_{02}
+K_{23} \}  +f_5(s), \\
T_6&= 2(K_{03}^2+K_{12}^2+K_{01}^2+K_{23}^2-L_{02}^2-
L_{13}^2)+f_6(s), \\ T_7&=
-4k_3K_{12}+(k_1+k_2)(K_{01}-L_{13})+(k_2-k_1)(L_{02}
+K_{23}).
\endalign
$$

For the Hamiltonian we have $H \thicksim T_6$.
\medskip
The commuting pairs of integrals can be organized as
follows:
\smallskip I. \underbar {\sl Two ignorable variables}.
$$
1.-\,\,(T_1, \,T_3),\qquad O(2,2) \supset E(1) \times E(1).
\tag 5.71
$$
\smallskip II. \underbar {\sl One ignorable variable}.
\smallskip
A). {\sl Subgroup type}.
$$
2.-\,\,(T_3, \,T_7^2),\qquad O(2,2) \supset E(1,1) \supset
O(1,1).
\tag 5.72
$$

B). {\sl Generic}.
$$
3.-\,\,(T_1, \,T_4 +aT_3). \tag 5.73
$$

III. \underbar {\sl No ignorable variables}.
\smallskip
A). {\sl Subgroup type}.
$$
4.-\,\,(T_3, \,T^2_7+aT_1+bT_2), \quad (a,\, b) \neq (0, \,
0). \tag 5.74
$$

B). {\sl Generic}.
$$
5.-\,\, (T_4+bT_3+aT_5, \,\, T_1+a^2T_2-aT_3), \quad a  \neq
0.
\tag 5.75$$
\bigskip
{\bf 10. The MANS $M_{11}$.}
\medskip
The reduced integrals are:
$$
\align
T_1&=\frac14(-K_{01}+L_{13}+L_{02}+K_{23})^2+f_1(s),\\
T_2&=\frac12(K_{01}-L_{13}+L_{02}+K_{23})^2+
\{ K_{12},\,-K_{01}+L_{13}+L_{02}+K_{23} \}+f_2(s), \\
T_3&=2(K_{03}^2+K_{12}^2+K_{01}^2+K_{23}^2-L_{13}^2-L_{02
}^2)+f_3(s) ,\\
T_4&=\frac12((L_{02}+K_{23})^2-(K_{01}-L_{13})^2) +f_4(s),
\tag 5.76 \\   T_5&=\frac12 \{
K_{03}-K_{12},\,-K_{01}+L_{13}+L_{02}+K_{23} \}  +f_5(s),
\\ T_6&= \frac12((K_{23}+L_{13})^2-(K_{01}-L_{02})^2-
\{ K_{03}+K_{12}, \,\, K_{01}-L_{13}+L_{02}+K_{23} \})+
f_6(s).
\endalign
$$

The integral $T_3$ is directly related to the Hamiltonian.
\medskip
The commuting pairs of integrals are:
\smallskip
II. \underbar{\sl One ignorable variable}.
\smallskip
A). {\sl Subgroup type}:
$$
1.-\,\,(T_1, \,T_4),\qquad O(2,2) \supset E(1,1).  \tag 5.77
$$
\smallskip
B). {\sl Generic}.
$$
2.-\,\,(T_1, \,T_5+aT_4). \tag 5.78
$$

III. \underbar{\sl No ignorable variables}.
\smallskip
A). {\sl Subgroup type}
$$
3.-\,\,(T_4, \,T_2). \tag 5.79
$$

B). {\sl Generic}.
$$
4.-\,\,(2abT_1+aT_2+bT_4+T_6, \,2bT_1+T_2+4aT_4+2T_5).  \tag
5.80
$$
\vfill\eject
%\bigskip \bigskip
\centerline {\bf 6.- CONCLUSIONS.}
\medskip
The results contained in this article can be situated in
three  complementary  research programs. They are: 1) The
classification of maximal abelian  subalgebras  of the
classical Lie algebras and their applications in physics. 2)
A systematic  search for integrable and ``superintegrable"
Hamiltonian systems in various  homogeneous spaces. 3) The
theory of the separation of variables in partial
differential equations, invariant under some Lie group of
local point  transformations.
\medskip
Physical applications that we have in mind can be either
direct or indirect  ones.  The integrability of the
considered systems makes it possible to calculate
eigenvalues and eigenfunctions of the quantum systems, as
well as trajectories in  the classical ones. The usefulness
of the above facts hinges on the question  whether the
deduced potentials can be associated with realistic physical
phenomena,  as was the case for
$O(2,1)$ potentials [13,
\dots, 17]. We mention that integrable  systems in spaces
with indefinite metric were studied by Kibler [28].
\medskip
A different and less direct application occurs in soliton
theory [29].  Indeed,  finite dimensional integrable systems
can sometimes be associated with infinite  dimensional ones.
Solutions of the finite dimensional  systems then provide
interesting special solutions  of the infinite dimensional
systems [30--32].     \bigskip \bigskip
{\bf Acknowledgements}
\medskip
The research of the first two authors is supported by
DGICYT de Espa\~{n}a (grants PB92--0255 and PB92--0197).
That of the third by grants from NSERC of Canada and FCAR
du Qu\'{e}bec. All three authors benefited from a NATO
International Collaborative Research Grant. Helpful
discussions with  J. Harnad  are acknowledged.
\bigskip
%\medskip

\centerline{\bf References.}
\smallskip

[1]. del Olmo, M.A., Rodr\'{\i}guez, M.A. and Winternitz,
P., J. Math.  Phys. {\bf 34}, 5118,

(1993).
\smallskip
[2]. Jacobson, N., {\sl Lie Algebras }, Wiley, New York,
(1962).
\smallskip
[3]. del Olmo, M.A., Rodr\'{\i}guez, M.A., Winternitz, P.
and Zassenhaus H.,  Lin. Alg.

and Applic. {\bf 135},79 (1990).
\smallskip
[4]. Wojciechowski, S., Phys. Lett. {\bf 95A}, 279 (1983);
{\bf 100A}, 471 (1984).
\smallskip
[5]. Evans, N.W., Phys. Rev. {\bf41}, 5666 (1990), Phys.
Lett. {\bf 147 A}, 483 (1990); J.

Math. Phys. {\bf 32}, 3369 (1991).
\smallskip
[6]. Fock, V.A., Z. Phys. {\bf 98}, 145 (1935).
\smallskip
[7]. Bargman, V., Z. Phys. {\bf 99}, 576 (1936).
\smallskip
[8]. Fri\v{s} I., Mandrosov V., Smorodinsky, Ya. A.,
Uhl\'{\i}\v{r}, M.  and Winternitz, P., Phys.

Lett. {\bf 16}, 354 (1965).
\smallskip
[9]. Makarov, A.A., Smorodinsky, Ya. A., Valiev, Kh.,  and
Winternitz, P.,

Nuovo Cim. {\bf A52}, 1061 (1967).
\smallskip
[10]. Kibler, M. and Winternitz, P., J. Phys. {\bf A 20},
4097 (1987); Phys. Lett.{\bf 147A},

338 (1990).
\smallskip
[11]. Kibler, M., Lamot, G.H., Winternitz, P., Int. J.
Quantum Chem. {\bf 43}, 625

(1992).
\smallskip
[12]. Moshinsky, M., {\sl The Harmonic Oscillator in  Modern
Physics: From Atoms to}

{\sl Quarks}, Gordon and Breach, New York, 1969.
\smallskip
[13]. P\"oschl,G. and Teller, E., Z. Phys. {\bf 83}, 143
(1933).
\smallskip
[14]. Alhassid, Y., Gursey, F. and Iachello, F., Phys. Rev.
Lett. {\bf 50}, 873 (1983);

Chem. Phys. Lett. {\bf 99}, 27 (1983).
\smallskip
[15]. Frank, A. and Wolf, K.B., Phys. Rev. Lett. {\bf 52},
1737 (1984).
\smallskip
[16]. Wulfman, C. and Levine, R.D., Chem. Phys. Lett.,  {\bf
60}, 372 (1979).
\smallskip
[17]. Wehrhahn, R.F., Smirnov, Yu.F., and Shirokov, A.M.,
J. Math. Phys. {\bf 33}, 2384

(1992).
\smallskip
[18]. Kobayashi, S. and Nomizu, K., {\sl  Foundations of
Differential Geometry I, II},

Wiley, New York, 1963 and 1969.
\smallskip
[19]. Eisenhart, L.P., Ann. Math. {\bf 35}, 284 (1934).
\smallskip
[20]. Miller, W. Jr., Patera, J. and Winternitz, P., J.
Math. Phys. {\bf 22}, 251 (1981).
\smallskip
[21]. Winternitz, P. and Fri\v{s} I., Sov. J. Nucl. Phys.
{\bf 1}, 636 (1965).
\smallskip
[22]. Winternitz, P., Luka\v{c}, I. and Smorodinsky, Ya. A.,
Sov. J. Nucl. Phys.  {\bf 7}, 139

(1968).
\smallskip
[23]. Miller, W. Jr., {\sl Symmetry and Separation of
Variables}, Addison Wesley, New

York, (1977).
\smallskip
[24]  Kalnins, E.G. and Miller, W. Jr., Proc. Roy. Soc.
Edinb. {\bf 79A},  227  (1977).
\smallskip
[25]. Kalnins, E.G. and Miller, W. Jr., SIAM J. Math. Anal.
{\bf 9}, 12 (1978).
\smallskip
[26]. Kalnins, E.G. and Miller, W. Jr., J. Diff. Geom.  {\bf
14}, 221 (1979).
\smallskip
[27]. Kalnins, E.G., SIAM J. Math. Anal. {\bf 6}, 340
(1975).
\smallskip
[28]. Kibler, M., in {\sl Group Theoretical Methods in
Physics}, Lecture Notes in Phys.

{\bf 318}, p. 238, Springer--Verlag, Berlin (1988).
\smallskip
[29]. Ablowitz, M.J. and Clarkson, P.A., {\sl Solitons,
Nonlinear  Evolution Equations}

{and \sl Inverse Scattering}, Cambridge University Press,
Cambridge  (1991).
\smallskip
[30]. Previato, E., Physica {\bf D18}, 312 (1986)., Duke
Math. J. {\bf 52}, 329 (1985).
\smallskip
[31]. Adams, M.R., Harnad, J. and Previato, E., Commun.
Math. Phys. {\bf 117}, 451

(1988).
\smallskip
[32]. Adams, M.R., Harnad, J. and Hurtubise, J., Commun.
Math. Phys. {\bf 134}, 555

(1990).
\vfill\eject
\pagebreak
%%%%%%%%%%%%%%%%%%%%%%%%%%%%%

\vskip 1cm

{\bf Table 1.}  MASAs of $su(2,2)$. The basis elements
$X_i$ are to be  identified  with the matrices in
eq.\,(2.3a--f). Which realization to  take is indicated by
the subscript of the metric $K_i$. The abbreviations used are
OD for an orthogonally decomposable MASA; OID means
orthogonally indecomposable, and D means decomposable, but
not orthogonally.
\bigskip

$$
\vbox{\offinterlineskip
\def\dd#1{\hfil#1\hfil}
\def\cc#1{\quad#1\hfil}
\def\hh#1{\hfil\quad #1\quad \hfil}
\def\ss#1{\hfil\  #1\ \hfil}
\hrule
\halign{&\vrule#&\strut\hh{$\displaystyle#$}
&\vrule#&\ss{$\displaystyle#$}
&\vrule#&\hh{$\displaystyle#$}&\vrule#
&\ss{$\displaystyle#$}\cr
height6pt&\omit&&\omit&&\omit&&\omit&\cr
&N_0&&\text{Type}&&\text{Basis:}\
Y_1,Y_2,Y_3&&\text{Realization}&\cr
height6pt&\omit&&\omit&&\omit&&\omit&
\cr
\noalign{\hrule} height6pt&\omit&&\omit&&\omit &&\omit&\cr
&M_1&&
\text{Cartan compact}&&X_1,X_2,X_3&&K_1&\cr
height4pt&\omit&&\omit&&\omit&&\omit&\cr & &&\text{OD}\
(1,0)+(0,1)+(1,0)+(0,1)&& && &\cr height4pt&
\omit&&\omit&&\omit&&\omit&\cr &M_2&&\text{Cartan
noncompact}&&X_1,X_1+2X_2+X_3,X_{15}&&K_1&\cr
height4pt&\omit&&\omit&&\omit&&\omit&\cr & &&\text{OD}\
(1,0)+(0,1)+ (1,1)&& && &\cr
height4pt&\omit&&\omit&&\omit&&\omit&\cr &M_3&&
\text{Cartan noncompact}&&X_1+2X_2+X_3,X_5, X_{15}&&K_1&\cr
height4pt&\omit&&\omit&&\omit&&\omit&\cr &  &&\text{OD}\
(1,1)+(1,1)&&  && &\cr
height4pt&\omit&&\omit&&\omit&&\omit&\cr &M_4&&\text{OD}\
(1,0)+(0,1)+(1,1)&&X_1,X_2,X_{14} &&K_2 &\cr
height4pt&\omit&&\omit&&\omit&&\omit&\cr  &M_5&&\text{OD}\
(1,1)+(1,1)&&X_2,X_5,X_{14} &&K_2 &\cr
height4pt&\omit&&\omit&&\omit&&
\omit&\cr &M_6&&\text{OD}\ (1,1)+ (1,1)&&X_3,X_4,X_{14}
&&K_3 &\cr height4pt&\omit&&\omit&&\omit&& \omit&\cr
&M_{7,8}&&\text{OD}\ (2,1)+(0,1)&&X_2+ 2X_3,X_4,X_{7}
&&K_{4,\varepsilon} &\cr
height4pt&\omit&&\omit&&\omit&&\omit&\cr &M_9&&\text{OID
\& D}&&X_5-X_7,X_8-X_9,X_{11} &&K_5 &\cr
height4pt&\omit&&\omit&&\omit&&
\omit&\cr &M_{10}&&\text{MANS (121)}&&X_5,X_7,X_{12} &&K_6
&\cr height4pt&\omit&&\omit&&\omit&&\omit&\cr
&M_{11}&&\text{MANS (121)}&&X_5+X_{13},X_7,X_{12} &&K_6
&\cr height4pt&\omit&&\omit&&\omit&&\omit&\cr
&M_{12}&&\text{MANS (202)}&&X_6,X_7,X_{12},X_{13} &&K_6 &\cr
height6pt&\omit&&\omit&&\omit&&\omit&\cr }\hrule}
$$

\vfill\eject
\pagebreak
%%%%%%%%%%%%%%%%%%%%%%%%%%%%%%%%%%%%%

{\bf APPENDIX 1.} Second order operators in the enveloping
algebra  commuting with MASAs. The operators are written in
the presentation corresponding to the metric $K_1$. A basis
of each MASA in this metric is also given. The notations
coincide with of those of Table I only for the Cartan
subalgebras $M_1, \, M_2$ and $M_3$.

\bigskip

{\bf MASA 1:}
$\{X_1,\,X_2,\,X_3\}$

$$\align T_1&=X_4^2+X_5^2\\ T_2&=X_6^2+X_7^2\\ T_3&=X_8^2
+X_9^2\\ T_4&=X_{10}^2+X_{11}^2\\ T_5&=X_{12}^2+X_{13}^2\\
T_6&=X_{14}^2+X_{15}^2
\endalign$$

\medskip

{\bf MASA 2:}
$\{X_1,\,X_1+2X_2+X_3,\,X_{15}\}$

$$\align T_1&=X_4^2+X_5^2\\ T_2&=-X_{14}^2+X_3^2\\
T_3&=X_6^2-X_8^2+X_7^2-X_9^2\\ T_4&=X_{10}^2-X_{12}^2
+X_{11}^2-X_{13}^2\\ T_5&=\{X_6,X_8\}+\{X_7,X_9\}\\
T_6&=\{X_{10},X_{12}\}+\{X_{11},X_{13}\}
\endalign$$

\medskip {\bf MASA 3:}
$\{X_1+2X_2+X_3,\,X_5,\,X_{15}\}$

$$\align T_1&=-X_4^2+X_1^2\\ T_2&=-X_{14}^2+X_3^2\\
T_3&=X_6^2-X_8^2-X_{10}^2+X_{12}^2+
      X_7^2-X_9^2-X_{11}^2+X_{13}^2\\
T_4&=\{X_6,X_8\}-\{X_{10},X_{12}\}+
      \{X_7,X_9\}-\{X_{11},X_{13}\}\\
T_5&=\{X_6,X_{10}\}-\{X_8,X_{12}\}+
      \{X_7,X_{11}\}-\{X_9,X_{13}\}\\
T_6&=\{X_6,X_{12}\}+\{X_8,X_{10}\}+
      \{X_7,X_{13}\}+\{X_9,X_{11}\}
\endalign$$

\pagebreak

{\bf MASA 4:}
$\{X_1,\,2X_2+X_3,\,X_3-X_{15}\}$

$$\align T_1&=X_4^2+X_5^2\\ T_2&=\frac{1}{2}(X_6-
X_8)^2+\frac{1}{2} (X_7-X_9)^2\\
T_3&=\frac{1}{2}(X_{10}-X_{12})^2+
      \frac{1}{2}(X_{11}-X_{13})^2\\ T_4&=
X_{14}^2-X_3^2+X_{15}^2\\ T_5&=X_6^2-X_8^2+ X_7^2-X_9^2\\
T_6&=X_{10}^2-X_{12}^2+X_{11}^2-X_{13}^2
\endalign$$

\medskip

{\bf MASA 5:}
$\{X_1+2X_2+X_3,\,X_5,\,X_3+X_{15}\}$

$$\align T_1&=-X_4^2+X_1^2\\ T_2&=X_{14}^2-
X_3^2+X_{15}^2\\ T_3&=\frac{1}{2}(X_6+X_8)^2-
      \frac{1}{2}(X_{10}+X_{12})^2+
      \frac{1}{2}(X_7+X_9)^2-
      \frac{1}{2}(X_{11}+X_{13})^2\\ T_4&=-
\frac{1}{2}\{X_6+X_8,X_{10}+ X_{12}\}-
       \frac{1}{2}\{X_7+X_9,X_{11}+X_{13}\}\\
T_5&=X_6^2-X_8^2-X_{10}^2+X_{12}^2+
      X_7^2-X_9^2-X_{11}^2+X_{13}^2\\ T_6&=
-\{X_6,X_{10}\}+\{X_8,X_{12}\}-
       \{X_7,X_{11}\}+\{X_9,X_{13}\}
\endalign$$

\medskip

{\bf MASA 6:}
$\{X_1+2X_2+X_3,\,X_1+X_5,\,X_3+X_{15}\}$

$$\align T_1&=\frac{1}{4}(X_6+X_8-X_{10}-X_{12})^2+
      \frac{1}{4}(X_7+X_9-X_{11}-X_{13})^2\\ T_2&=
X_4^2-X_1^2+X_5^2\\ T_3&=X_{14}^2-X_3^2+X_{15}^2\\
T_4&=\frac{1}{2}(X_6-X_{10})^2-
      \frac{1}{2}(X_8-X_{12})^2+
      \frac{1}{2}(X_7-X_{11})^2-
      \frac{1}{2}(X_9-X_{13})^2\\ T_5&=
\frac{1}{2}(X_6+X_8)^2-
      \frac{1}{2}(X_{10}+X_{12})^2+
      \frac{1}{2}(X_7+X_9)^2-
      \frac{1}{2}(X_{11}+X_{13})^2\\ T_6&
=X_6^2-X_8^2-X_{10}^2+X_{12}^2+
      X_7^2-X_9^2-X_{11}^2+X_{13}^2
\endalign$$

\vfil\eject
\pagebreak

{\bf MASA 7:}
$\{X_1+2X_2+3X_3,\,X_2-X_{11},\,X_5-X_7\}$

$$\align T_1&=\frac{1}{2}(X_{12}+X_{14})^2+
      \frac{1}{2}(X_{13}+X_{15})^2\\ T_2&=
\frac{1}{2}(X_4-X_6)^2+
      \frac{1}{2}\{X_1-X_3,X_2-X_{11}\}\\
T_3&=X_8^2-X_{12}^2+X_{14}^2+
      X_9^2-X_{13}^2+X_{15}^2\\ T_4&=4(
X_4^2-X_6^2+X_{10}^2)+
      4(X_5^2-X_7^2+X_{11}^2)-\\
    &5X_1^2-4X_2^2+3X_3^2-
      2\{X_1,X_2\}+\{X_1,X_3\}+
      2\{X_2,X_3\}\\ T_5&=\frac{1}{\sqrt{2}}
\{X_8,X_{12}+X_{14}\}+
      \frac{1}{\sqrt{2}}\{X_9,X_{13}+X_{15}\}\\
T_6&=\sqrt{2}\{X_{10}, X_4-X_6\}+
      \frac{1}{\sqrt{2}}\{X_1-X_3,X_5-X_7\}+\\
    &\sqrt{2}\{X_2-X_{11},X_5+X_7\}
\endalign$$

\medskip {\bf MASA 8:}
\{$X_1+2X_2-X_3,\,X_1+X_2+X_3-X_9,\,X_5+X_{13}\}$

$$\align T_1&=\frac{1}{2}(X_6+X_{14})^2+
      \frac{1}{2}(X_7-X_{15})^2\\ T_2&=
\frac{1}{2}(X_4-X_{12})^2
     -\frac{1}{2}\{X_1-X_3,X_1+X_2+X_3-X_9\}\\
T_3&=-X_6^2+X_{10}^2+ X_{14}^2
      -X_7^2+X_{11}^2+X_{15}^2\\ T_4&=4(X_4^2+X_8^2-
X_{12}^2)+
      4(X_5^2+X_9^2-X_{13}^2)-
      5X_1^2-4X_2^2-5X_3^2-\\
    &2\{X_1,X_2\}-3\{X_1,X_3\}-
      6\{X_2,X_3\}\\ T_5&=\frac{1}{\sqrt{2}}\{X_{10},
X_6+X_{14}\}+
      \frac{1}{\sqrt{2}}\{X_{11},X_7-X_{15}\}\\ T_6&=
\sqrt{2}\{X_8,X_4 -X_{12}\}+
      \frac{1}{\sqrt{2}}\{X_1-X_3,X_5+X_{13}\}-\\
    &\sqrt{2}\{X_1+X_2+X_3-X_9,X_5-X_{13}\}
\endalign$$

\medskip

{\bf MASA 9:}
$\{X_9+X_{11},\,X_1-X_3+X_5-X_{15},
\,X_7+X_9-X_{11}-X_{13}\}$

$$\align T_1&=\frac{1}{4}(X_6+X_8-X_{10}-X_{12})^2-
      \frac{1}{2}\{X_1+X_5,X_3+X_{15}\}\\
T_2&=(X_4+X_{14})^2-(X_6+X_{12})^2-
      (X_1-X_3)^2+(X_5-X_{15})^2-\\
    &(X_7-X_{13})^2+
      (X_9-X_{11})^2
\endalign$$
$$\align T_3&=(X_4-X_{14})^2-(X_6-X_{12})^2+2(X_8^2+
X_{10}^2)
     -2(X_1+X_2)^2-\\
    &2(X_2+X_3)^2-2\{X_1,X_3\}
     +(X_5+X_{15})^2-(X_7+X_{13})^2+\\
    &(X_9+X_{11})^2\\ T_4&=\frac{1}{2}\{X_4-X_{14},X_6
+X_8-X_{10}- X_{12}\}+\\
    &\frac{1}{2}\{X_1+X_5,X_7-X_9-X_{11}+X_{13}\}+\\
    &\frac{1}{2}\{X_3+X_{15},X_7+X_9+X_{11}+X_{13}\}\\
T_5&=X_{10}^2- X_8^2-
      \frac{1}{2}\{X_6-X_{12},X_8+X_{10}\}
     +(X_1+X_3)^2+\\
      &\{X_1+X_3,X_2\}+
      \frac{1}{2}\{X_1+2X_2+X_3,X_5+X_{15}\}\\ T_6&=
-\{X_4+X_{14},X_6+ X_{12}\}
     -\{X_5-X_{15},X_7-X_{13}\}+\\
    &\{X_1-X_3,X_9-X_{11}\}
\endalign$$

\medskip

{\bf MASA 10:}
$\{X_1+X_2+X_3+X_9,\,X_5-X_{13},\, X_7+X_{15}\}$

$$\align T_1&=\frac{1}{4}(X_4-X_6+X_{12}+X_{14})^2
     +\frac{1}{2}\{X_1+X_2+X_3+X_9,X_2-X_{11}\}\\
T_2&=\frac{1}{4}(X_4+X_6+X_{12}-X_{14})^2
     +\frac{1}{2}\{X_1+X_2+X_3+X_9,X_2+X_{11}\}\\
T_3&=-\frac{1}{2}(X_4+X_{12})^2+
       \frac{1}{2}(X_6-X_{14})^2\\
   &+X_1^2-X_3^2+\frac{1}{2}\{X_1-X_3,X_2+X_9\}\\
T_4&=-\frac{1}{2}\{X_4-X_6+X_{12}+X_{14},X_8+X_{10}\}\\
   &-\frac{1}{2}\{X_1+2X_2+X_3+X_9-X_{11},X_5+X_{15}\}\\
   &+\frac{1}{2}\{X_1+X_3+X_9+X_{11},X_7-X_{13}\}\\
T_5&=\frac{1}{2}\{X_4+X_6+X_{12}-X_{14},X_8-X_{10}\}\\
&+\frac{1}{2}\{X_1+2X_2+X_3+X_9+X_{11},X_5-X_{15}\}\\
&+\frac{1}{2}\{X_1+X_3+X_9-X_{11},X_7+X_{13}\}\\
T_6&=2(X_4^2+X_8^2+X_{10}^2+X_{14}^2-X_6^2-X_{12}^2)
     -3X_1^2-4X_2^2-3X_3^2-\\
    &2\{X_1,X_2\}-\{X_1,X_3\}-2\{X_2,X_3\}+\\
    &2(X_5^2-X_7^2+X_9^2+X_{11}^2-X_{13}^2+X_{15}^2)
\endalign$$

\medskip {\bf MASA 11:}
$\{X_1+X_2+X_3+X_9,\,X_5+X_7-X_{13}+X_{15},\, 2X_5-2X_{13}
-X_2-X_{11}\}$

$$\align T_1&=\frac{1}{4}(X_4+X_6+X_{12}-X_{14})^2+\\
  &\frac{1}{2}\{X_1+X_2+X_3+X_9,X_5-X_7-X_{13}-X_{15}\}\\
T_2&=\frac{1}{2}(X_4-X_6+X_{12}+X_{14})^2
     +\{X_{10},X_4+X_6+X_{12}-X_{14}\}+\\
    &\frac{1}{2}(X_5-X_7-X_{13}-X_{15})^2
     +\{X_2-X_{11},X_1+X_2+X_3+X_9\}+\\
    &\frac{1}{2}\{X_7+X_{15},-X_1+2X_2+X_3+2X_{11}\}+\\
    &\frac{1}{2}\{X_5-X_{13},-X_1-2X_2+X_3-2X_{11}\}\\
T_3&=2(X_4^2+X_8^2+X_{10}^2+X_{14}^2-X_6^2-X_{12}^2)
     -3X_1^2-4X_2^2-3X_3^2-\\
     &2\{X_1,X_2\}
     -\{X_1,X_3\}-2\{X_2,X_3\}+\\
    &2(X_5^2-X_7^2+X_9^2+X_{11}^2-X_{13}^2+X_{15}^2)\\
T_4&=-\frac{1}{2}(X_4+X_{12})^2+\frac{1}{2}(X_6-X_{14})^2
     +X_1^2-X_3^2+\frac{1}{2}\{X_1,X_2\}-\\
    &\frac{1}{2}\{X_2,X_3\}
     +\frac{1}{2}\{X_9,X_1-X_3\}
     -\frac{1}{2}(X_5-X_{13})^2
     +\frac{1}{2}(X_7+X_{15})^2\\
T_5&=\frac{1}{2}\{X_4+X_6+X_{12}-X_{14}),X_8-X_{10}\}+\\
    &\frac{1}{2}\{X_1+2X_2+X_3+X_9+X_{11},X_5-X_{15}\}+\\
    &\frac{1}{2}\{X_1+X_3+X_9-X_{11},X_7+X_{13}\}\\
T_6&=-\frac{1}{2}(X_4+X_6)^2+
       \frac{1}{2}(X_{12}-X_{14})^2-\\
    &\frac{1}{2}\{X_4-X_6+X_{12}+X_{14},X_8+X_{10}\}-\\
&\frac{1}{2}\{X_1+2X_2+X_3+X_9-X_{11},X_5+X_{15}\}+\\
    &\frac{1}{2}\{X_1+X_3+X_9+X_{11},X_7-X_{13}\}-\\
      &\frac{1}{2}\{X_1,X_2\}+\frac{1}{2}\{X_2,X_3\}
       -\frac{1}{2}(X_5+X_7)^2
       -\frac{1}{2}\{X_{11},X_1-X_3\}+\\
      &\frac{1}{2}(X_{13}-X_{15})^2
\endalign$$

\vfill\eject
\pagebreak
%%%%%%%%%%%%%%%%%%%%%%%%%%%%%%%%%

{\eightpoint{\bf Appendix 2.}  Commutation tables of  second
order   elements commuting with each MASA. The operators $A$,
$B$, $C$, $D$ are linearly  independent third order elements
in the enveloping algebra and are  different in each case.}
\smallskip

%\comment
\hbox{
\hbox{\vbox{\noindent MASA 1:
\medskip
\vbox{\offinterlineskip
\def\dd#1{\hfil$\;$#1$\;$}
\hrule
\halign{
\vrule#&
\dd{$\displaystyle#$}&
\vrule#&
\dd{$\displaystyle#$}&
\dd{$\displaystyle#$}&
\dd{$\displaystyle#$}&
\dd{$\displaystyle#$}&
\dd{$\displaystyle#$}&
\dd{$\displaystyle#$}&
\vrule#\cr height2pt&\omit&&\omit&\omit&\omit&
\omit&\omit&\omit&\cr &&&T_1&T_2&T_3&T_4&T_5&T_6&\cr
height2pt&\omit&&\omit&\omit&\omit&\omit&\omit& \omit&\cr
\noalign{\hrule} height2pt&\omit&&\omit&\omit&
\omit&\omit&\omit&\omit&\cr &T_1&& &A&B&A&B&0&\cr
height2pt&\omit&&\omit&\omit&\omit&\omit&\omit&
\omit&\cr &T_2&& & &C&A&0&-C&\cr
height2pt&\omit&&\omit&\omit&\omit&\omit&\omit&
\omit&\cr &T_3&& & & &0&-B&-C&\cr
height2pt&\omit&&\omit&\omit&\omit&\omit&\omit&\omit&\cr
&T_4&& & & & &D&D&\cr
height2pt&\omit&&\omit&\omit&\omit&\omit&\omit&\omit&\cr
&T_5&& & & & & &D&\cr
height2pt&\omit&&\omit&\omit&\omit&\omit&\omit&\omit&\cr
&T_6&& & & & & & &\cr
height2pt&\omit&&\omit&\omit&\omit&\omit&\omit&\omit&\cr
}\hrule}}}\hskip -6cm
\hbox{\vbox{\noindent MASA 2:
\smallskip
%\smallskip
%\smallskip
\vbox{\offinterlineskip
\def\dd#1{\hfil$\;$#1$\;$}
\hrule
\halign{
\vrule#&
\dd{$\displaystyle#$}&
\vrule#&
\dd{$\displaystyle#$}&
\dd{$\displaystyle#$}&
\dd{$\displaystyle#$}&
\dd{$\displaystyle#$}&
\dd{$\displaystyle#$}&
\dd{$\displaystyle#$}&
\vrule#\cr height2pt&\omit&&\omit&\omit&\omit&
\omit&\omit&\omit&\cr &&&T_1&T_2&T_3&T_4&T_5&T_6&\cr
height2pt&\omit&&\omit&\omit&\omit&\omit&\omit&\omit&\cr
\noalign{\hrule}\cr height2pt&\omit&&\omit&
\omit&\omit&\omit&\omit&
\omit&\cr &T_1&& &0&A&A&B&B&\cr
height2pt&\omit&&\omit&\omit&\omit&\omit&\omit& \omit&\cr
&T_2&& & &0&0&C&D&\cr
height2pt&\omit&&\omit&\omit&\omit&\omit&\omit&\omit&\cr
&T_3&& & & &A&-C&B&\cr
height2pt&\omit&&\omit&\omit&\omit&\omit&\omit&\omit&\cr
&T_4&& & & & &-B&D&\cr
height2pt&\omit&&\omit&\omit&\omit&\omit&\omit&\omit&\cr
&T_5&&  & & & & &-A&\cr
height2pt&\omit&&\omit&\omit&\omit&\omit&\omit&\omit&\cr
&T_6&& & & & & & &\cr
height2pt&\omit&&\omit&\omit&\omit&\omit&\omit&\omit&\cr
}\hrule}}}}

\smallskip
%\smallskip
%\smallskip

\hbox{
\hbox{\vbox{\noindent MASA 3:
\smallskip
%\smallskip
%\smallskip
\vbox{\offinterlineskip
\def\dd#1{\hfil$\;$#1$\;$}
\def\cc#1{\hfil$\;$#1$\;\hfil$}
\hrule
\halign{
\vrule#&
\dd{$\displaystyle#$}&
\vrule#&
\dd{$\displaystyle#$}&
\dd{$\displaystyle#$}&
\dd{$\displaystyle#$}&
\dd{$\displaystyle#$}&
\cc{$\displaystyle#$}&
\cc{$\displaystyle#$}&
\vrule#\cr height2pt&\omit&&\omit&\omit&\omit&\omit&
\omit&\omit&\cr &&&T_1&T_2&T_3&T_4&T_5&T_6&\cr
height2pt&\omit&&\omit&\omit&\omit&\omit&\omit&\omit&\cr
\noalign{\hrule} height2pt&\omit&&\omit&\omit&\omit&
\omit&\omit&\omit&\cr &T_1&& &0&0&0&A&B&\cr
height2pt&\omit&&\omit&\omit&\omit&\omit&\omit&
\omit&\cr &T_2&& & &0&C&0&D&\cr
height2pt&\omit&&\omit&\omit&\omit&\omit&\omit&\omit&\cr
&T_3&& & & &-C&-A&-B-D&\cr
height2pt&\omit&&\omit&\omit&\omit&\omit&\omit&\omit&\cr
&T_4&& & & & &-B+D&A&\cr
height2pt&\omit&&\omit&\omit&\omit&\omit&\omit&\omit&\cr
&T_5&& & & & & &C&\cr
height2pt&\omit&&\omit&\omit&\omit&\omit&\omit&\omit&\cr
&T_6&& & & & & & &\cr
height2pt&\omit&&\omit&\omit&\omit&\omit&\omit&\omit&\cr
}\hrule}}}\hskip -6cm
\hbox{\vbox{\noindent MASA 4:
\smallskip
%\smallskip
%\smallskip
\vbox{\offinterlineskip
\def\dd#1{\hfil$\;$#1$\;$}
\def\cc#1{\hfil$\;$#1$\;\hfil$}
\hrule
\halign{
\vrule#&
\dd{$\displaystyle#$}&
\vrule#&
\dd{$\displaystyle#$}&
\dd{$\displaystyle#$}&
\dd{$\displaystyle#$}&
\dd{$\displaystyle#$}&
\dd{$\displaystyle#$}&
\dd{$\displaystyle#$}&
\vrule#\cr height2pt&\omit&&\omit&\omit&\omit&\omit&
\omit&\omit&\cr &&&T_1&T_2&T_3&T_4&T_5&T_6&\cr
height2pt&\omit&&\omit&\omit&\omit&\omit&\omit&\omit&\cr
\noalign{\hrule} height2pt&\omit&&\omit&\omit&\omit&
\omit&\omit&\omit&\cr &T_1&& &A&A&0&B&B&\cr
height2pt&\omit&&\omit&\omit&\omit&\omit&\omit&
\omit&\cr &T_2&& & &0&C&C&A&\cr
height2pt&\omit&&\omit&\omit&\omit&\omit&\omit&\omit&\cr
&T_3&& & & &D&-A&-D&\cr
height2pt&\omit&&\omit&\omit&\omit&\omit&\omit&\omit&\cr
&T_4&& & & & &0&0&\cr
height2pt&\omit&&\omit&\omit&\omit&\omit&\omit&\omit&\cr
&T_5&& & & & & &B&\cr
height2pt&\omit&&\omit&\omit&\omit&\omit&\omit&\omit&\cr
&T_6&& & & & & & &\cr
height2pt&\omit&&\omit&\omit&\omit&\omit&\omit&\omit&\cr
}\hrule}}}}

\smallskip
%\smallskip
%\smallskip

\hbox{
\hbox{\vbox{\noindent MASA 5:
\smallskip
%\smallskip
%\smallskip
\vbox{\offinterlineskip
\def\dd#1{\hfil$\;$#1$\;$}
\def\cc#1{\hfil$\;$#1$\;\hfil$}
\hrule
\halign{
\vrule#&
\dd{$\displaystyle#$}&
\vrule#&
\dd{$\displaystyle#$}&
\dd{$\displaystyle#$}&
\dd{$\displaystyle#$}&
\dd{$\displaystyle#$}&
\cc{$\displaystyle#$}&
\cc{$\displaystyle#$}&
\vrule#\cr height2pt&\omit&&\omit&\omit&\omit&\omit&
\omit&\omit&\cr &&&T_1&T_2&T_3&T_4&T_5&T_6&\cr
height2pt&\omit&&\omit&\omit&\omit&\omit&\omit&\omit&\cr
\noalign{\hrule} height2pt&\omit&&\omit&\omit&\omit&
\omit&\omit&\omit&\cr &T_1&& &0&0&A&0&B&\cr
height2pt&\omit&&\omit&\omit&\omit&\omit&\omit&
\omit&\cr &T_2&& & &C&D&0&0&\cr
height2pt&\omit&&\omit&\omit&\omit&\omit&\omit&\omit&\cr
&T_3&& & & &0&-C&-A-D&\cr
height2pt&\omit&&\omit&\omit&\omit&\omit&\omit&\omit&\cr
&T_4&& & & & &A-D&C&\cr
height2pt&\omit&&\omit&\omit&\omit&\omit&\omit&\omit&\cr
&T_5&& & & & & &-B&\cr
height2pt&\omit&&\omit&\omit&\omit&\omit&\omit&\omit&\cr
&T_6&& & & & & & &\cr
height2pt&\omit&&\omit&\omit&\omit&\omit&\omit&\omit&\cr
}\hrule}}}\hskip -6cm
\hbox{\vbox{\noindent MASA 6:
\smallskip
%\smallskip
%\smallskip
\vbox{\offinterlineskip
\def\dd#1{\hfil$\;$#1$\;$}
\def\cc#1{\hfil$\;$#1$\;\hfil$}
\hrule
\halign{
\vrule#&
\dd{$\displaystyle#$}&
\vrule#&
\dd{$\displaystyle#$}&
\dd{$\displaystyle#$}&
\dd{$\displaystyle#$}&
\dd{$\displaystyle#$}&
\cc{$\displaystyle#$}&
\cc{$\displaystyle#$}&
\vrule#\cr height2pt&\omit&&\omit&\omit&\omit&\omit&\omit
&\omit&\cr &&&T_1&T_2&T_3&T_4&T_5&T_6&\cr
height2pt&\omit&&\omit&\omit&\omit&\omit&\omit&\omit&\cr
\noalign{\hrule} height2pt&\omit&&\omit&\omit&\omit&\omit
&\omit&\omit&\cr &T_1&& &A&B&0&0&A+B&\cr
height2pt&\omit&&\omit&\omit&\omit&\omit&
\omit&\omit&\cr &T_2&& & &0&C&0&0&\cr
height2pt&\omit&&\omit&\omit&\omit&\omit&\omit&\omit&\cr
&T_3&& & & &0&D&0&\cr
height2pt&\omit&&\omit&\omit&\omit&\omit&\omit&\omit&\cr
&T_4&& & & & &A-B&-C&\cr
height2pt&\omit&&\omit&\omit&\omit&\omit&\omit&\omit&\cr
&T_5&& & & & & &-D&\cr
height2pt&\omit&&\omit&\omit&\omit&\omit&\omit&\omit&\cr
&T_6&& & & & & & &\cr
height2pt&\omit&&\omit&\omit&\omit&\omit&\omit&\omit&\cr
}\hrule}}}}

\smallskip
%\smallskip
%\smallskip

\hbox{
\hbox{\vbox{\noindent MASA 7,8:
\smallskip
%\smallskip
%\smallskip
\vbox{\offinterlineskip
\def\dd#1{\hfil$\;$#1$\;$}
\def\cc#1{\hfil$\;$#1$\;\hfil$}
\hrule
\halign{
\vrule#&
\dd{$\displaystyle#$}&
\vrule#&
\dd{$\displaystyle#$}&
\dd{$\displaystyle#$}&
\dd{$\displaystyle#$}&
\dd{$\displaystyle#$}&
\dd{$\displaystyle#$}&
\dd{$\displaystyle#$}&
\vrule#\cr height2pt&\omit&&\omit&\omit&\omit&\omit&
\omit&\omit&\cr &&&T_1&T_2&T_3&T_4&T_5&T_6&\cr
height2pt&\omit&&\omit&\omit&\omit&\omit&\omit&\omit&\cr
\noalign{\hrule} height2pt&\omit&&\omit&\omit&\omit&\omit&
\omit&\omit&\cr &T_1&& &0&A&-4A&B&-2B&\cr
height2pt&\omit&&\omit&\omit&\omit&\omit&\omit&\omit&\cr
&T_2&& & & 0&0&B&C&\cr
height2pt&\omit&&\omit&\omit&\omit&\omit&\omit&\omit&\cr
&T_3&& & & &0&D&0&\cr
height2pt&\omit&&\omit&\omit&\omit&\omit&\omit&\omit&\cr
&T_4&& & & & &-4D&0&\cr
height2pt&\omit&&\omit&\omit&\omit&\omit&\omit&\omit&\cr
&T_5&& & & & & &-2A&\cr
height2pt&\omit&&\omit&\omit&\omit&\omit&\omit&\omit&\cr
&T_6&& & & & & & &\cr
height2pt&\omit&&\omit&\omit&\omit&\omit&\omit&\omit&\cr
}\hrule}}}\hskip -6cm
\hbox{\vbox{\noindent MASA 9:
\smallskip
%\smallskip
%\smallskip
\vbox{\offinterlineskip
\def\dd#1{\hfil$\;$#1$\;$}
\def\cc#1{\hfil$\;$#1$\;\hfil$}
\hrule
\halign{
\vrule#&
\dd{$\displaystyle#$}&
\vrule#&
\dd{$\displaystyle#$}&
\dd{$\displaystyle#$}&
\dd{$\displaystyle#$}&
\dd{$\displaystyle#$}&
\dd{$\displaystyle#$}&
\dd{$\displaystyle#$}&
\vrule#\cr height2pt&\omit&&\omit&\omit&\omit&\omit&
\omit&\omit&\cr &&&T_1&T_2&T_3&T_4&T_5&T_6&\cr
height2pt&\omit&&\omit&\omit&\omit&\omit&\omit&\omit&\cr
\noalign{\hrule} height2pt&\omit&&\omit&\omit&\omit&\omit
&\omit&\omit&\cr &T_1&& &A&-A&0&0&B&\cr
height2pt&\omit&&\omit&\omit&\omit&\omit&\omit&
\omit&\cr &T_2&& & &0&C&D&0&\cr
height2pt&\omit&&\omit&\omit&\omit&\omit&\omit&\omit&\cr
&T_3&& & & &-C&-D&0&\cr
height2pt&\omit&&\omit&\omit&\omit&\omit&\omit&\omit&\cr
&T_4&& & & & &B&-D&\cr
height2pt&\omit&&\omit&\omit&\omit&\omit&\omit&\omit&\cr
&T_5&& & & & & &C&\cr
height2pt&\omit&&\omit&\omit&\omit&\omit&\omit&\omit&\cr
&T_6&& & & & & & &\cr
height2pt&\omit&&\omit&\omit&\omit&\omit&\omit&\omit&\cr
}\hrule}}}}

\smallskip
%\smallskip
%\smallskip
%\endcomment
\hbox{
\hbox{\vbox{\noindent MASA 10:
\smallskip
%\smallskip
%\smallskip
\vbox{\offinterlineskip
\def\dd#1{\hfil$\;$#1$\;$}
\def\cc#1{\hfil$\;$#1$\;\hfil$}
\hrule
\halign{
\vrule#&
\dd{$\displaystyle#$}&
\vrule#&
\dd{$\displaystyle#$}&
\dd{$\displaystyle#$}&
\dd{$\displaystyle#$}&
\dd{$\displaystyle#$}&
\dd{$\displaystyle#$}&
\dd{$\displaystyle#$}&
\dd{$\displaystyle#$}&
\vrule#\cr height2pt&\omit&&\omit&\omit&\omit&\omit&
\omit&\omit&\omit&\cr &&&T_1&T_2&T_3&T_4&T_5&T_6&T_7&\cr
height2pt&\omit&&\omit&\omit&\omit&\omit&\omit&\omit&
\omit&\cr
\noalign{\hrule} height2pt&\omit&&\omit&\omit&\omit&
\omit&\omit&\omit&
\omit&\cr &T_1&& &A&0&0&B&0&E&\cr
height2pt&\omit&&\omit&\omit&\omit&\omit&\omit&\omit&
\omit&\cr &T_2&& & &0&C&0&0&F&\cr
height2pt&\omit&&\omit&\omit&\omit&\omit&\omit&\omit&
\omit&\cr &T_3&& & & &B&C&0&0&\cr
height2pt&\omit&&\omit&\omit&\omit&\omit&\omit&\omit&
\omit&\cr &T_4&& & & & &D&0&G&\cr
height2pt&\omit&&\omit&\omit&\omit&\omit&\omit&\omit&
\omit&\cr &T_5&& & & & & & &H&\cr
height2pt&\omit&&\omit&\omit&\omit&\omit&\omit&\omit&
\omit&\cr &T_6&&  & & & & & &0&\cr
height2pt&\omit&&\omit&\omit&\omit&\omit&\omit&\omit&
\omit&\cr &T_7&& & & & & & &&\cr
height2pt&\omit&&\omit&\omit&\omit&\omit&\omit&\omit&
\omit&\cr }\hrule}}}\hskip -6cm
\hbox{\vbox{\noindent MASA 11:
\smallskip
%\smallskip
%\smallskip
\vbox{\offinterlineskip
\def\dd#1{\hfil$\;$#1$\;$}
\def\cc#1{\hfil$\;$#1$\;\hfil$}
\hrule
\halign{
\vrule#&
\dd{$\displaystyle#$}&
\vrule#&
\dd{$\displaystyle#$}&
\dd{$\displaystyle#$}&
\dd{$\displaystyle#$}&
\dd{$\displaystyle#$}&
\dd{$\displaystyle#$}&
\dd{$\displaystyle#$}&
\vrule#\cr height2pt&\omit&&\omit&\omit&\omit&\omit&
\omit&\omit&\cr &&&T_1&T_2&T_3&T_4&T_5&T_6&\cr
height2pt&\omit&&\omit&\omit&\omit&\omit&\omit&\omit&\cr
\noalign{\hrule} height2pt&\omit&&\omit&\omit&\omit&\omit&
\omit&\omit&\cr &T_1&& &A&0&0&0&B&\cr
height2pt&\omit&&\omit&\omit&\omit&\omit&\omit&
\omit&\cr &T_2&& & &0&0&C&D&\cr
height2pt&\omit&&\omit&\omit&\omit&\omit&\omit&\omit&\cr
&T_3&& & & &0&0&0&\cr
height2pt&\omit&&\omit&\omit&\omit&\omit&\omit&\omit&\cr
&T_4&& & & & &B&C/2&\cr
height2pt&\omit&&\omit&\omit&\omit&\omit&\omit&\omit&
\cr
&T_5&& & & & & &-D/2&\cr
height2pt&\omit&&\omit&\omit&\omit&\omit&\omit&\omit&\cr
&T_6&& & & &  & & &\cr
height2pt&\omit&&\omit&\omit&\omit&\omit&\omit&\omit&\cr
}\hrule}}}}

\vfill\eject
\end